\documentclass[aps,prb,showpacs,floatfix,twocolumn,superscriptaddress,byrevtex]{revtex4-1}
\pdfoutput=1
\usepackage{amsmath}
\usepackage{amssymb}
\usepackage{amstext}
\usepackage{amsopn}
\usepackage{amsfonts}
\usepackage{amsxtra}
\usepackage{color}
\usepackage{dcolumn}
\usepackage{graphicx}
\usepackage{hyperref}
\usepackage{bm}
\begin{document}

%
%
\title{Optical properties and electronic structure of the non-metallic metal FeCrAs}
\author{A. Akrap}
\altaffiliation{Present address: University of Geneva, CH-1211 Geneva 4, Switzerland}
\affiliation{Condensed Matter Physics and Materials Science Department,
  Brookhaven National Laboratory, Upton, New York 11973, USA}%
\author{Y. M. Dai}
\affiliation{Condensed Matter Physics and Materials Science Department,
  Brookhaven National Laboratory, Upton, New York 11973, USA}%
\author{W. Wu}
\affiliation{Department of Physics, University of Toronto -- 60 St. George Street,
  Toronto, Ontario M5S 1A7, Canada}
\author{S. R. Julian}
\affiliation{Department of Physics, University of Toronto -- 60 St. George Street,
  Toronto, Ontario M5S 1A7, Canada}
\affiliation{Canadian Institute for Advanced Research, 180 Dundas St. West, Suite 1400,
  Toronto, M5G 1Z8, Canada}
\author{C. C. Homes}
\email{homes@bnl.gov}
\affiliation{Condensed Matter Physics and Materials Science Department,
  Brookhaven National Laboratory, Upton, New York 11973, USA}%
\date{\today}

%
%
\begin{abstract}
The complex optical properties of a single crystal of hexagonal FeCrAs ($T_N \simeq 125$\,K)
have been determined above and below $T_N$ over a wide frequency range in the planes (along
the {\em b} axis), and along the perpendicular ({\em c} axis) direction.  At room temperature,
the optical conductivity $\sigma_1(\omega)$ has an anisotropic metallic character.
The electronic band structure reveals two bands crossing the Fermi level, allowing the
optical properties to be described by two free-carrier (Drude) contributions consisting of
a strong, broad component and a weak, narrow term that describes the increase in $\sigma_1(\omega)$
below $\simeq 15$~meV.
The dc-resistivity of FeCrAs is ``non-metallic'', meaning that it rises in power-law fashion with decreasing temperature,
without any signature of a transport gap.  In the analysis of the optical conductivity, the scattering
rates for both Drude contributions track the dc-resistivity quite well, leading us to conclude
that the non-metallic resistivity of FeCrAs is primarily due to a scattering rate that increases
with decreasing temperature, rather than the loss of free carriers.  The power law $\sigma_1(\omega)
\propto \omega^{-0.6}$ is observed in the near-infrared region and as $T\rightarrow T_N$ spectral
weight is transferred from low to high energy ($\gtrsim 0.6$~eV); these effects may be explained
by either the two-Drude model or Hund's coupling.
We also find that a low-frequency in-plane phonon mode decreases in frequency for $T < T_N$,
suggesting the possibility of spin-phonon coupling.

\end{abstract}
%
%
\maketitle
%
%

%
%
\section{Introduction}
\begin{figure}[tb]
\includegraphics[width=0.85\columnwidth]{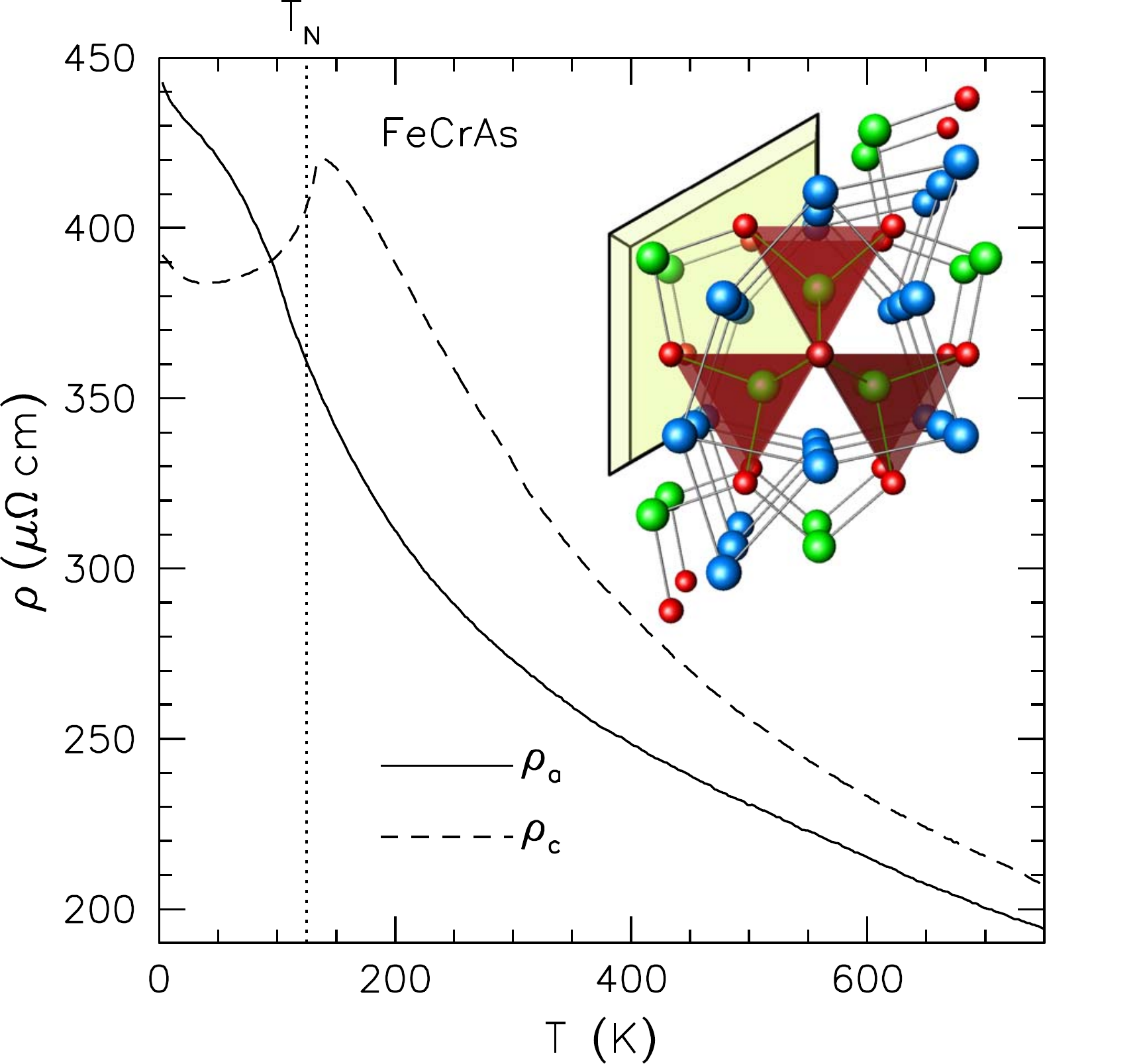}
\caption{The resistivity of FeCrAs along the {\em a} axis (solid line) and
{\em c} axis (dashed line); along the {\em c} axis there is an anomaly just
above $T_N \simeq 125$~K (dotted line).
The inset shows the projection of the extended unit cell of FeCrAs
along the crystallographic {\em c} axis (onto the hexagonal {\em a-b}
plane).  The unit cell is indicated by the tinted region.  The polyhedra
indicate the As (red) coordination around the Fe atoms (green), with the
Cr atoms (blue) also forming corner-sharing triangles.}
\label{fig:resis}
\end{figure}

FeCrAs is a rare example of a ``non-metallic metal'', a material whose resistivity rises with
decreasing temperature, but with power-law temperature dependence, as opposed to the exponential
rise that signals an insulating gap in the excitation spectrum.  Other non-metallic metals are underdoped
cuprates,\cite{ando95,dobrosavljevic12} some heavy-fermion systems,\cite{maple95} materials on
the border of Anderson localization,\cite{mott70} dilute Kondo systems, quasicrystalline
metals,\cite{rosenbaum07} and the parent compounds of some iron-pnictide superconductors
above their magnetic/structural ordering temperatures.\cite{stewart11}  In most of these
cases, non-metallic behavior is not understood.
FeCrAs is quite an extreme example of non-metallic behavior:\cite{wu09} from above 800~K to
below 80~mK, the resistivity in the hexagonal (\emph{a-b}) plane rises monotonically,
from $\sim200\,\mu\Omega\,$cm at 800~K to $\sim450\,\mu\Omega\,$cm at 100~mK, shown
in Fig.~\ref{fig:resis}, values that are in the metallic range suggesting a high carrier
density. The \emph{c}-axis resistivity is similar (Fig.~\ref{fig:resis}), except for a peak
just above an antiferromagnetic ordering temperature $T_N\simeq 125$~K that briefly interrupts
the rising resistivity.

Moreover, as $T\rightarrow 0$ K, the resistivity does not saturate. Rather, it continues to
rise with a sublinear power-law. This is in contrast, for example, to dilute Kondo systems in
which the resistivity saturates as $\rho(T) = \rho_0 - A T^2$, behavior that Nozi\`{e}res
explained as indicative of the Fermi-liquid ground state that is expected at temperatures low
enough that internal degrees of freedom are locked.\cite{nozieres74}  Also, the thermodynamic
properties of FeCrAs at low temperature {\em are} Fermi-liquid-like, but quite enhanced, with
a large $T$-linear specific heat $C(T)/T \sim 30$ mJ/mole\,K$^2$ and a correspondingly large
Pauli-like magnetic susceptibility $\chi(T)$, leading to a Wilson ratio of $R_W \simeq 4$,
as opposed to the expected value of 2 for a Kondo system.\cite{wu09}  This combination of
non-metallic, non-Fermi-liquid transport with Fermi-liquid thermodynamic properties is very
unusual.

The hexagonal crystal structure of FeCrAs, shown in the inset of Fig.~\ref{fig:resis}, has a
number of triangular motifs that suggest magnetic frustration. The Fe sublattice is a
triangular lattice of trimers, while the Cr sites form a distorted Kagome lattice of
corner-sharing triangles. Moreover, the antiferromagnetic order is typical of frustrated
metallic magnets, being an incommensurate spin density wave with no detectable ordered
moment on the Fe site and the Cr ordered moment varying between 0.6 and 2.2~$\mu_B$
throughout the magnetic unit cell, which has an in-plane lattice parameter that is
three times larger than the paramagnetic unit cell.\cite{swainson10}
%
%
In a recent study,\cite{tafti13} pressure was used to suppress the magnetic ordering transition,
and it was demonstrated that formation of a spin-density-wave gap does not play a decisive role in the
anomalous transport properties of this material.  It is, however, quite possible that magnetic
frustration is important, as recently suggested by Rau {\em et al.},\cite{rau11} who put forward
a theory that the anomalous behavior of FeCrAs arises from a ``hidden spin liquid'' on the iron
sublattice.  In this model the non-Fermi liquid, non-metallic behavior arises from strong charge
fluctuations on the Fe sublattice due to proximity to a metal-insulator transition.

%
%

An alternative scenario that has been proposed for FeCrAs\cite{nevidomskyy09,craco10} is that
it is a ``Hund's Metal'',\cite{yin11,haule08,haule09} a model that grew out of efforts to
explain the electronic properties, including incoherent transport, in superconducting iron-pnictides
and iron-chalcogenides. Dynamical mean field theory calculations for these systems suggest
that local Hund's rule correlations in the occupancies of the $d$-orbitals, combined with
destructive interference between direct and indirect hopping paths, can result in a subset
of the {\em d}-orbitals becoming strongly correlated, or even localized. The result is large
local moments coupled to more itinerant electrons.\cite{yin11} As pointed out in
Ref.~\onlinecite{nevidomskyy09}, treating these slowly fluctuating large moments in a Kondo
limit is equivalent to the underscreened Kondo model, in which the large magnitude of the
spin leads to exponential suppression of the Kondo temperature.\cite{nevidomskyy09,okada73}
This model predicts a rising resistivity down to very low temperature, and although there
should eventually be a crossover to Fermi liquid behavior, this may be at unobservably
low temperatures if the fluctuating moment is large enough. Calculations have mostly focused
on iron sites tetrahedrally coordinated by pnictogens (as in FeCrAs) or chalcogens; however,
it should be kept in mind that the large magnetic moment in FeCrAs is on the
pyramidally-coordinated Cr site. This model predicts that the optical masses should
be many times larger than the band masses, and that the Wilson ratio should also be
large,\cite{haule08,haule09} as observed in FeCrAs. The prediction for the optical
conductivity is that it will show a fractional power law dependence on frequency,
$\sigma_1(\omega) \propto \omega^{-\alpha}$ with $0 < \alpha < 1$, in the near- and
mid-infrared regions.\cite{yin12}

%
%
The optical conductivity contains a wealth of information about the transport and
electrodynamics of a material.\cite{dressel-book}  For example, the multiband
$A$Fe$_2$As$_2$ (``122'') materials,\cite{graser10} where $A=$ Ba, Ca or Sr, are
the parent compounds for a number of iron-based superconductors.\cite{johnston10}
These materials are poorly-metallic and undergo structural and magnetic transitions at
$T_s \simeq 130-200$~K, below which their dc conductivity increases
substantially;\cite{rotter08,tanatar09} angle-resolved photoemission spectroscopy
(ARPES) reveals that below $T_s$ the Fermi surface is partially gapped\cite{richard10}
in response to the formation of a spin density wave, resulting in a decrease in the
density of states at the Fermi level.  In spite of this Fermi-surface reconstruction
these compounds remain metallic and show decreasing resistivity at low temperature.
This behavior is reflected in the optical conductivity; below $T_s$ there is a transfer
of spectral weight (the area under the conductivity curve) from low to high frequency,
indicating the opening of a transport gap and consistent with the partial gapping of
the Fermi surface observed in ARPES.  The decrease in the low-frequency spectral weight
also signals the loss of free carriers; however, this loss is compensated by the dramatic
decrease in the scattering rate below $T_s$ with the net result being a resistivity that
decreases at low temperature.\cite{hu08,akrap09}
On the other hand, the presence of strong correlations or strong disorder, such as in
the quasicrystal Al$_{63.5}$Cu$_{24.5}$Fe$_{12}$, can lead to a dramatically different
conductivity spectrum, where the Drude contribution is almost entirely
absent.\cite{homes91,timusk13}

In this paper, we present the first temperature-dependent polarized infrared spectra for
single crystal FeCrAs.  Electronic structure calculations were also performed which show
that this material is expected to be metallic, with two energy bands crossing the Fermi
level.  We have described this multiband conductor using two free-carrier (Drude)
contributions; the first is a strong, broad term, while the second is a weak, narrow component
that is necessary to describe the low-frequency behavior.  The mid-infrared response is
described by two bound excitations.  An anisotropic response is observed in both the free-carrier
response and low-frequency bound excitation.  The conductivity in the mid and near-infrared
regions is assumed to arise from bound (interband) excitations which might mask the carrier
response in the incoherent regime; however, we note that a $\sigma_1(\omega) \propto
\omega^{-0.6}$ is observed in this energy range which may lend support to the view that
this material is a Hund's metal.  The coherent Drude components, however, show a scattering
rate whose temperature dependence follows the dc-resistivity quite closely, demonstrating
that the anomalous non-metallic resistivity of FeCrAs is primarily due to anomalous
temperature dependence of the scattering rate, rather than localization of charge carriers.
In addition to the large-scale electronic structure, sharp features in the conductivity are
observed at low frequency that are attributed to the normally-active infrared lattice modes.
One in-plane lattice mode softens below $T_N$, suggesting the presence of spin-phonon
coupling.

%
%
\begin{figure}[tb]
\includegraphics[width=0.90\columnwidth]{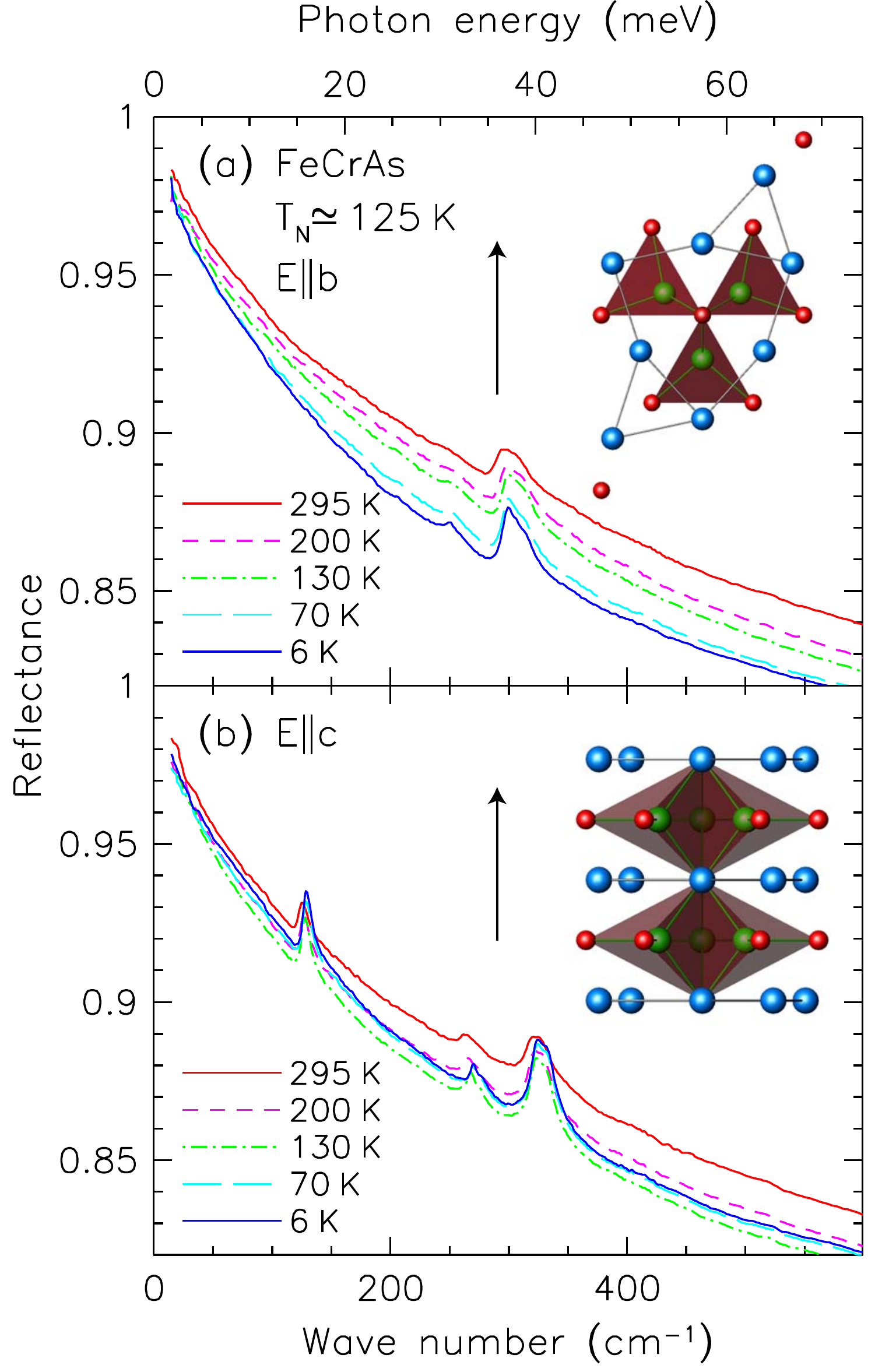}
\caption{The temperature dependence of the reflectance of a single crystal
of FeCrAs for (a) light polarized along the {\em b} axis direction in the
infrared region.  Inset: The orientation of the electric field along the
crystallographic {\em b} axis.  The Fe, Cr and As atoms are green, blue and
red, respectively.
(b) The reflectance for light polarized along the {\em c} axis in the infrared
region. Inset:  The orientation of the electric field along the crystallographic
{\em c} axis.}%
\label{fig:reflec}
\end{figure}
%

%
%
\section{Experimental}
Crystals were grown from a stoichiometric melt in an alumina crucible within
a sealed quartz tube.  The material was melted twice, and then annealed at
900$^\circ$C for 150 hours.  Sample quality in FeCrAs is revealed by the
sharpness of the resistive transition near $T_N$, the value of $T_N$, and the
onset temperature and magnitude of glassy behavior, in the form  of a difference
between field-cooled and zero-field-cooled magnetic susceptibility.
The crystals used in these measurements were from our highest quality
batch,\cite{wu09} in which $T_N \sim 125$~K in susceptibility measurements,
there is a sharp cusp in the {\em c}-axis resistivity about 8~K above $T_N$
(Fig.~\ref{fig:resis}), and glassy behavior is barely visible and
only sets in below 10~K.  Further details of crystal growth and
characterization can be found in Ref.~\onlinecite{wu11}.
The crystal for the infrared measurements was oriented using Laue-diffraction,
and then cut using spark-erosion to expose a flat {\em b-c} surface approximately
$4.6\,{\rm mm}\times 1.8\,{\rm mm}$, where the longer dimension is along the
{\em c} axis.  The sample was subsequently polished using fine diamond paste,
revealing a flat, mirror-like surface.

%
%
The reflectance has been measured at a near-normal angle of incidence over
a wide frequency range ($\simeq 2$~meV to 5~eV) for light
polarized along the {\em b}- and {\em c}-axis directions using an overfilling
technique and referenced using an {\em in situ} evaporation method.\cite{homes93}
The low-frequency reflectance is shown in Figs.~\ref{fig:reflec}(a) and \ref{fig:reflec}(b)
for the {\em b}- and {\em c}-axis directions, respectively.  The orientation of the
electric field along the crystallographic direction is shown in the insets.
The temperature dependence of the reflectance was measured up to 1~eV, above
which the optical properties were assumed to be temperature-independent.
The polarization dependence of the reflectance was examined up to 3~eV.
The low-frequency reflectance for light polarized along the {\em b} axis,
shown in Fig.~\ref{fig:reflec}(a), displays a metallic character, but
decreases as the temperature is lowered, in agreement with the increasing
resistivity at low temperature.  The reflectance along the {\em c} axis,
shown in Fig.~\ref{fig:reflec}(b), is also metallic and decreasing with
temperature, below $T_N$ it increases slightly, again in agreement with
transport measurements.   The sharp features in the reflectance
are attributed to the normally-active infrared lattice vibrations (which
will be discussed below).

%
%
\begin{table}[b]
\caption{The experimental and theoretical lattice constants and atomic fractional
coordinates$^a$ for the optimized structure of FeCrAs in the hexagonal $P\bar{6}2m$
space group.  Note that $c/a$ has been fixed to the experimental value.}
\begin{ruledtabular}
\begin{tabular}{ccc}
 & Experiment$^b$    & Theory (GGA) \\
 $a$(\AA )           & 6.096  & 5.989 \\
 $c$(\AA )           & 3.651  & 3.609 \\
 Fe $(x0{1\over 2})$ & 0.2505 & 0.2522 \\
 Cr $(x00)$          & 0.5925 & 0.5877 \\
\end{tabular}
\end{ruledtabular}
\footnotetext[1] {The positions of the As atoms at $(000)$ and $({1\over 3}{2\over 3}{1\over 2})$
are fixed.}
\footnotetext[2] {Refs.~\onlinecite{hollan66,nyland72,geurin77}.  }
\label{tab:gga}
%
%
\end{table}

While the reflectance is a useful quantity, it depends on both the real and imaginary
parts of the dielectric function, $\tilde\epsilon(\omega) = \epsilon_1 + i\epsilon_2$,
and can therefore be difficult to interpret.  In order to work with more intuitive
quantities, the complex dielectric function, and in particular the real part of
the complex conductivity, is calculated from a Kramers-Kronig analysis of the
reflectance.\cite{dressel-book} Because this approach requires that the reflectance
be determined over the entire frequency interval, extrapolations must be supplied
for $\omega\rightarrow 0,\infty$.  At low frequency the material is a poor metal
and so the Hagen-Rubens form for the reflectance is employed, $R(\omega) =
1-a\sqrt{\omega}$, where $a$ is chosen to match the data at the lowest-measured
frequency point.  While other extrapolations have been considered, e.g.~a marginal-Fermi
liquid form\cite{littlewood91} with $R(\omega) \propto 1 - a\,\omega$, we have determined
that the results of the Kramers-Kronig analysis are largely insensitive to the details
of the low-frequency extrapolation.  Above the highest-measured frequency the reflectance
is assumed to be constant up to $7.5\times 10^4$~cm$^{-1}$, above which a free-electron
gas asymptotic reflectance extrapolation $R(\omega) \propto 1/\omega^4$ is
assumed.\cite{wooten}

%
%
\begin{figure}[t]
\includegraphics[width=0.99\columnwidth]{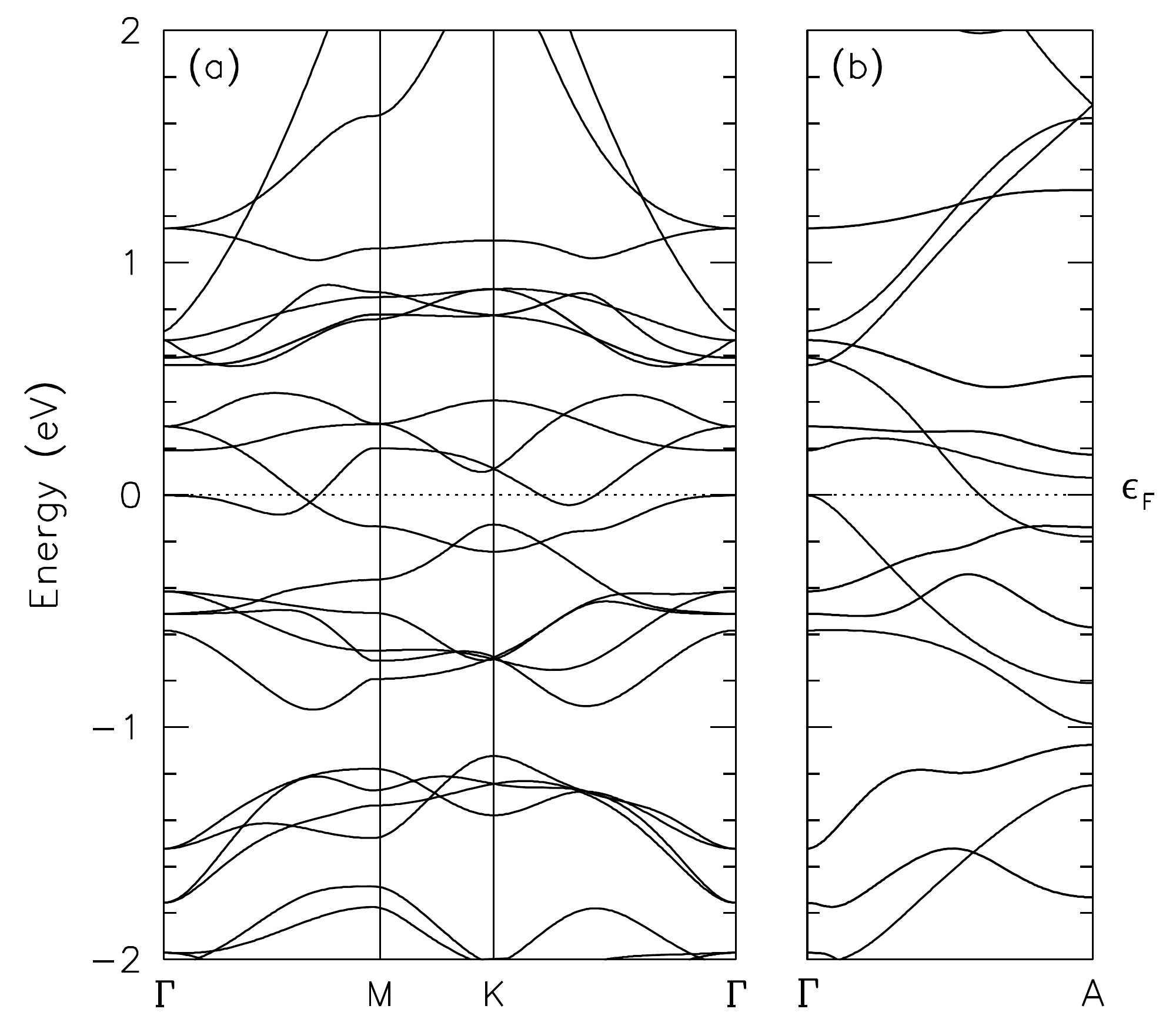}
\caption{The electronic band structure of hexagonal FeCrAs close to
the Fermi level ($\epsilon_F$) along several high-symmetry paths
of a hexagonal unit cell in the (a) $k_x$, $k_y$ planes and along the
(b) $k_z$ direction, respectively. The system is metallic, with two
bands crossing $\epsilon_F$.}
\label{fig:bands}
\end{figure}
%

%
%
%
\section{First principles methods and results}
Electronic structure calculations of FeCrAs were performed within the generalized
gradient approximation (GGA) using the full-potential linearized augmented plane-wave
(FP-LAPW) method\cite{singh} with local-orbital extensions\cite{singh91} in the
WIEN2k implementation.\cite{wien2k}  In this instance it was assumed that the
material is paramagnetic and that the spins do not interact.  An examination of
different Monkhorst-Pack {\em k}-point meshes indicated that a $4\times{4}\times{7}$
mesh and $R_{mt}k_{max}=8.5$ was sufficient for good energy convergence.
Initially, the volume of the unit cell was optimized with respect to the total
energy; during this process, the $c/a$ ratio was fixed to the experimentally-determined
value;\cite{hollan66,nyland72,geurin77} further geometric optimization was achieved
by relaxing the atomic fractional coordinates within the unit cell for the Fe and
Cr atoms until the total force was typically less than 1~mRy/a.u.  The results of
optimization of the volume and fractional atomic coordinates are summarized in
Table~\ref{tab:gga}.

\begin{figure}[t]
\includegraphics[width=0.90\columnwidth]{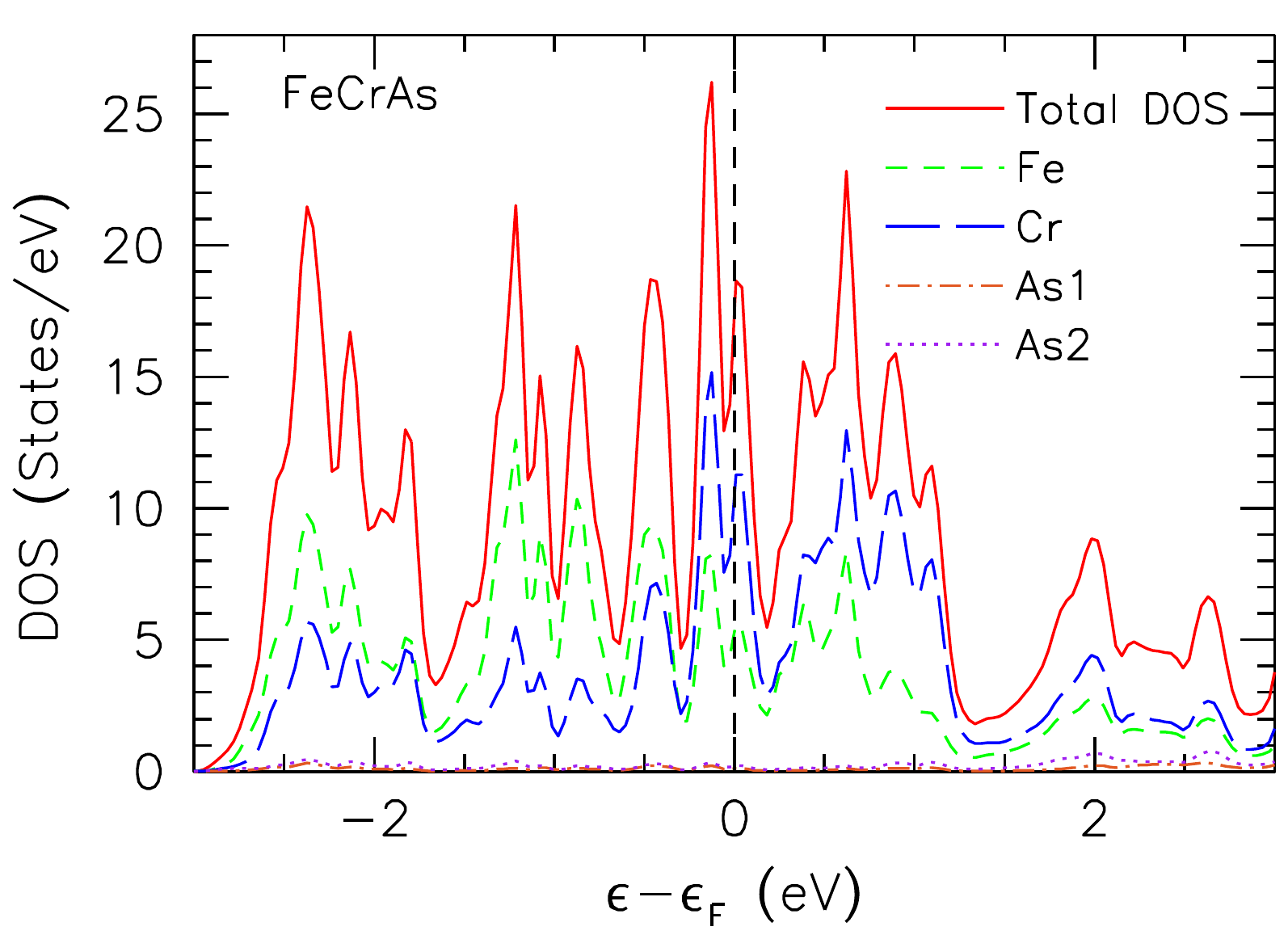}
\caption{The density of states for FeCrAs close to the Fermi level.  The
non-zero density of states at $\epsilon_F$ indicates a metallic
character, with the primary contributions coming from the Cr and Fe atoms,
while the As atoms play little role.}
\label{fig:dos}
\end{figure}

The electronic band structure along the high-symmetry directions for a hexagonal
unit cell near the Fermi level are shown in Fig.~\ref{fig:bands}.   Two bands cross the
Fermi level ($\epsilon_F$), indicating a metallic system, in agreement with experiment.
The orbital character indicates that the bands near $\epsilon_F$ are primarily
Fe and Cr $3d$ in nature, while the contribution from the As atoms is negligible.
This result is reflected in the density of states (DOS) at the Fermi level, shown in
Fig.~\ref{fig:dos}.  The total density at $\epsilon_F$ is reasonably high, with the main
contribution coming from Cr, followed by Fe; the As atoms contribute little or nothing
to the DOS at $\epsilon_F$, in agreement with a previous calculation of the density of
states.\cite{ishida96}

%
%

\begin{figure}[tb]
\includegraphics[width=0.65\columnwidth]{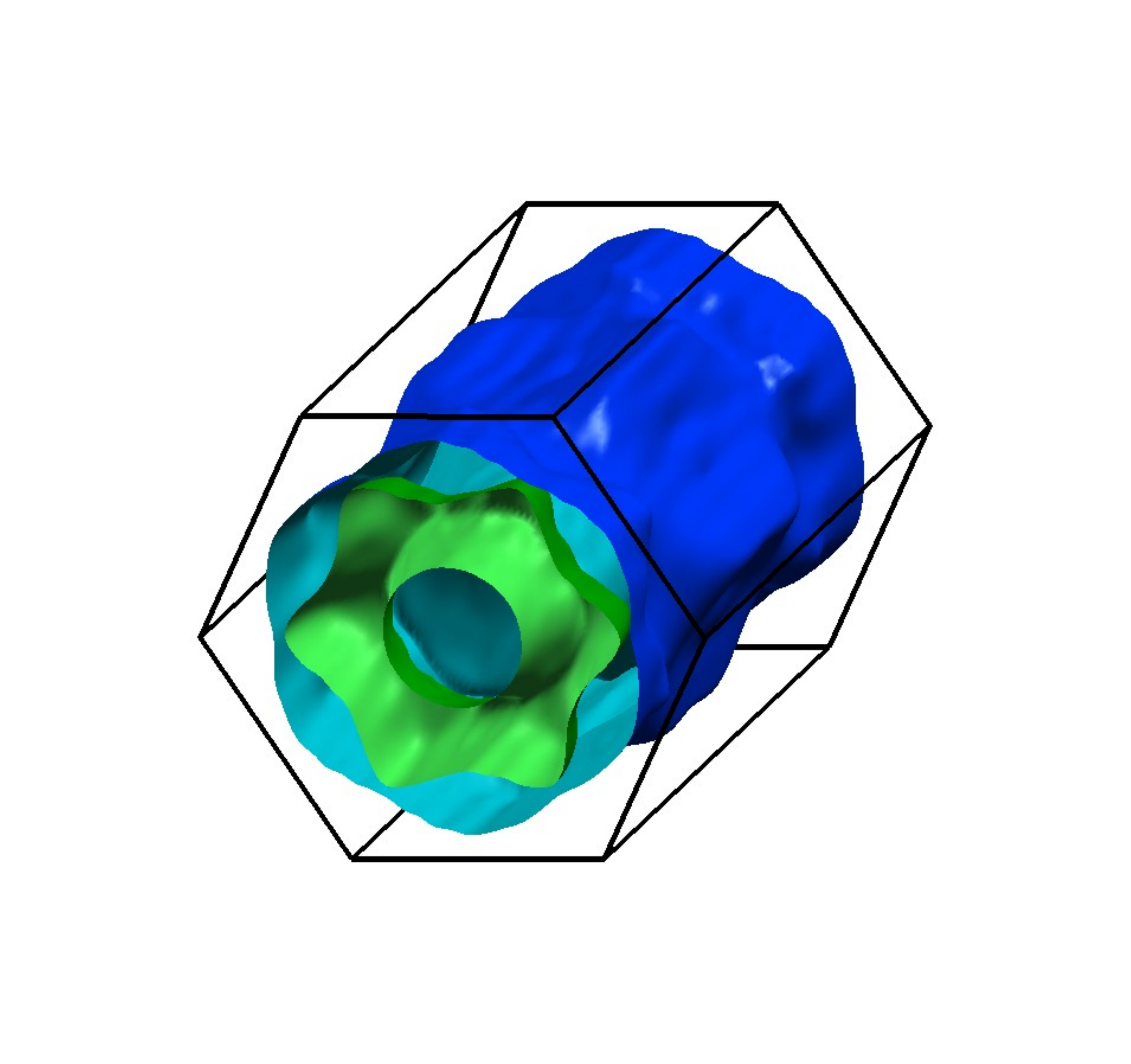}
\caption{The Fermi surfaces in the Brillouin zone of hexagonal FeCrAs.}
\label{fig:surface}
\end{figure}

The Fermi surfaces composed of the two bands that cross $\epsilon_F$ is shown
in Fig.~\ref{fig:surface}.  The outer sheet consists of a warped cylinder plus
an inner pocket centered on $\Gamma$, along the $k_z$ projection, while the
inner sheet may be described as a distorted torus centered on A.

%
%
%
\section{Results and Discussion}
\subsection{\boldmath $E\parallel{b}$ \unboldmath}
\begin{figure}[tb]
\includegraphics[width=0.85\columnwidth]{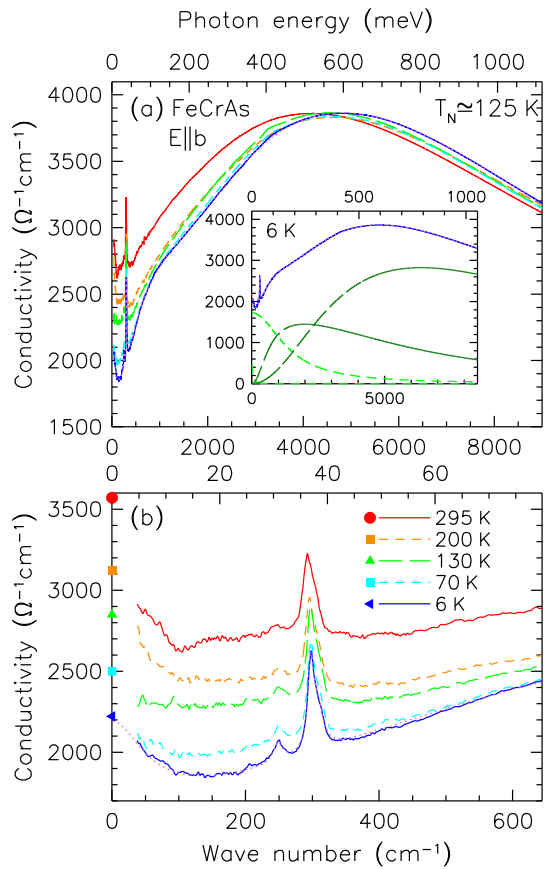}
\caption{(a) The temperature dependence of the real part of the optical conductivity
of FeCrAs for light polarized along the {\em b} axis over a wide frequency
range; note the shoulder that becomes visible at $\simeq 0.12$~eV at or below
$T_N$.  The dotted line indicates the model fit to the optical conductivity at 6~K.
Inset: the decomposition of the fit to the conductivity at 6~K into two Drude components
(dashed lines) and two Lorentzian oscillators (dash-dot lines).
(b) The temperature dependence of the real part of the optical conductivity shown
below 80~meV; the dotted line is the result of the fit to the data at 6~K.
The symbols at the origin indicate the transport values for $\sigma_{dc}$ along
this crystalline axis.}
\label{fig:sigb}
\end{figure}

The temperature dependence of the real part of the optical conductivity for
light polarized along the {\em b} axis is shown over a wide frequency range in
Fig.~\ref{fig:sigb}(a).  The low-frequency conductivity is shown in more detail
in Fig.~\ref{fig:sigb}(b) where the symbols at the origin denote the values for
$\sigma_{dc}$ determined from transport measurements (Fig.~\ref{fig:resis}).
At room temperature the optical conductivity may be described as a poor metal with
the conductivity displaying a broad maximum at $\sim 0.5$~eV.
A standard approach to fitting the optical conductivity in which contributions both
free carriers and a bound excitation are considered is the Drude-Lorentz model for
the dielectric function.  While the peculiar shape of the optical conductivity below
about 15~meV might suggest a renormalization process and therefore the generalized
Drude model,\cite{allen77,puchkov96} the presence of two bands at the Fermi level
precludes its use and instead suggests that contributions from two different types
of free carriers should be considered, the so-called ``two-Drude'' model,\cite{wu10}
\begin{equation}
  \tilde\epsilon(\omega) = \epsilon_\infty - \sum_j{{\omega_{p,D;j}^2}\over{\omega^2+i\omega/\tau_{D,j}}}
    + \sum_k {{\Omega_k^2}\over{\omega_k^2 - \omega^2 - i\omega\gamma_k}},
\end{equation}
where $\epsilon_\infty$ is the real part of the dielectric function at high
frequency, $\omega_{p,D;j}^2 = 4\pi n_je^2/m_j^\ast$ and $1/\tau_{D,j}$ are the square
of the plasma frequency and scattering rate for the delocalized (Drude) carriers in
the $j$th band, respectively, $n_j$ and $m^\ast_j$ are the carrier concentration and
effective mass; $\omega_k$, $\gamma_k$ and $\Omega_k$ are the position, width, and
strength of the $k$th vibration.  The complex conductivity is $\tilde\sigma(\omega)
= \sigma_1 +i\sigma_2 = -i\omega [\tilde\epsilon(\omega) - \epsilon_\infty ]/60$ (in
units of $\Omega^{-1}$cm$^{-1}$).

%
%
%

%
%
The fits to the optical conductivity are complicated somewhat by the broad,
overlapping nature of the bound excitations and the free-carrier components,
indeed the strong contribution from the bound excitations prevents us from
analyzing the low-frequency conductivity in terms of the Hund's metal
power-law frequency dependence discussed in the introduction.\cite{yin12}
In an effort to deal with this issue, the optical conductivity has first been
fit using a non-linear least-squares technique at 6~K, where in addition to the
features observed at room temperature, a strong shoulder is observed in the optical
conductivity at $\simeq 0.12$~eV.  The most reliable fit is achieved by using two Drude
components and two Lorentz oscillators; the fit over a wide frequency range is shown
by the dotted line in Fig.~\ref{fig:sigb}(a).
%
%
%
\begin{table}[tb]
\caption{The temperature dependence along the {\em b} and {\em c} axes of the
Drude scattering rates $1/\tau_{D,j}$, as well as the frequencies and damping
strong mid-infrared electronic absorptions $\omega_{L,k}$ and $\gamma_{L,k}$.
All units are in cm$^{-1}$ unless otherwise indicated.$^a$}
\begin{ruledtabular}
\begin{tabular}{c c c cc cc}
  \multicolumn{7}{c}{$E\parallel{b}$} \\
  T (K) &
  $1/\tau_{D,1}$ & $1/\tau_{D,2}$ & $\omega_{L,1}$ & $\gamma_{L,1}$ & $\omega_{L,2}$ & $\gamma_{L,2}$ \\
\cline{1-7}
%
%
    295 &   900 &  $\sim 35$ &    1850 & 4850 &  6590 & 15630  \\
    200 &  1000 &  $\sim 45$ &    2050 & 5340 &  6760 & 15560  \\
    130 &  1080 &  $\sim 55$ &    2080 & 5740 &  6530 & 15610  \\
     70 &  1270 &  $\sim 60$ &    2040 & 6480 &  6350 & 15680  \\
      6 &  1370 &  $\sim 65$ &    1950 & 6670 &  6360 & 15380  \\

  & & & & & & \\
%
%
 \multicolumn{7}{c}{$E\parallel{c}$} \\
  T (K) &
  $1/\tau_{D,1}$ & $1/\tau_{D,2}$ & $\omega_{L,1}$ & $\gamma_{L,1}$ & $\omega_{L,2}$ & $\gamma_{L,2}$ \\
\cline{1-7}
    295 & 350 &  $\sim 60$ & 1340 & 4420 &  6590 & 15120   \\
    200 & 400 &  $\sim 80$ & 1490 & 4540 &  6640 & 15030  \\
    130 & 440 &  $\sim 90$ & 1580 & 4650 &  6700 & 15070 \\
     70 & 410 &  $\sim 70$ & 1530 & 4710 &  6830 & 15330  \\
      6 & 400 &  $\sim 80$ & 1510 & 4690 &  6760 & 15110  \\
\end{tabular}
\end{ruledtabular}
\footnotetext[1] {For $E\parallel{b}$: $\omega_{p,D;1} \simeq 11\,740$~cm$^{-1}$,
  $\omega_{p,D;2} \approx 1460$~cm$^{-1}$, $\Omega_{L,1} \simeq 24\,130$ ~cm$^{-1}$
  and $\Omega_{L,2} \simeq 50\,980$~cm$^{-1}$, are constant.  For $E\parallel{c}$: $\omega_{p,D;1}
  \simeq 7090$~cm$^{-1}$, $\omega_{p,D;2} \simeq 1450$~cm$^{-1}$, $\Omega_{L,1}
  \simeq 26\,640$ ~cm$^{-1}$ and $\Omega_{L,2} \simeq 48\,610$~cm$^{-1}$ are also fixed.}
\label{tab:drude}
%
%
\end{table}
%
%
The Drude components consist of a strong broad component with $\omega_{p,D;1} \simeq
11\,740$~cm$^{-1}$ and $1/\tau_{D,1}\simeq 1370$~cm$^{-1}$, and a much weaker narrow
component with $\omega_{p,D;2}\simeq 1460$~cm$^{-1}$ and $1/\tau_{D,2}\simeq 35$~cm$^{-1}$,
while the parameters for the two Lorentz oscillators are $\omega_{L,1}\simeq 1950$~cm$^{-1}$,
$\gamma_{L,1}\simeq 6670$~cm$^{-1}$ and $\Omega_{L,1}\simeq 24\,130$~cm$^{-1}$ and
$\omega_{L,2}\simeq 6360$~cm$^{-1}$, $\gamma_{L,2}\simeq 15\,380$~cm$^{-1}$ and
$\Omega_{L,2}\simeq 50\,980$~cm$^{-1}$ (Table~\ref{tab:drude}).
The individual contributions to the optical conductivity are shown in the inset
of Fig.~\ref{fig:sigb}(a); it is immediately apparent that the
overdamped Lorentz oscillators show a considerable amount of overlap with
the broad Drude component.  We note that while the strong Drude component can reproduce
most of the low-frequency conductivity, the Drude value for $\sigma_1(\omega\rightarrow 0)
\equiv \omega_{p,D}^2\tau_D/60 \simeq 1680$~$\Omega^{-1}$cm$^{-1}$ is well below the
measured transport value of $\sigma_{dc} \simeq 2200$~$\Omega^{-1}$cm$^{-1}$.
The second, weak Drude component accurately reproduces the slight upturn in the optical
conductivity observed a low frequency, and $1/\tau_{D,2}$ has been further fit to ensure
that $\sigma_1(\omega\rightarrow 0) \simeq \sigma_{dc}$; this extrapolation is shown by the
dotted line in Fig.~\ref{fig:sigb}(b), where the low-frequency conductivity is described
quite well.  Comparing the two Lorentz oscillators, we note that the majority of the
spectral weight is associated with the high frequency oscillator; $(\Omega_{L,1} /
\Omega_{L,2})^2 \simeq 4.5$.
In order to perform similar fits for the remaining temperatures, a constraint has
been introduced on the fitting process by assuming that the Drude plasma frequencies
and the strengths of the Lorentzian oscillators do not change; only the Drude
scattering rates and the frequencies and widths of the Lorentz oscillators have
been allowed to vary.  The results of the fits using this convention are tabulated
in Table~\ref{tab:drude}.

%
%

%
%
In addition to the strong electronic absorptions, there are several sharp features
shown in more detail in Fig.~\ref{fig:sigb}(b) that are resolved as three separate
vibrations.  At 295~K they are centered at $\simeq 246, 292$ and 302~cm$^{-1}$.
The fitted parameters for these modes are listed in Table~\ref{tab:irmodes} at 295,
130 and 6~K.  FeCrAs crystallizes in the hexagonal $P\bar{6}2m$ space
group,\cite{hollan66,nyland72,geurin77} so the irreducible vibrational representation is
\begin{equation}
  \Gamma_{vib} = 2A_1^\prime + 2A_2^\prime + 6E^\prime + A_1^{\prime\prime} +
  3A_2^{\prime\prime} + 2E^{\prime\prime}. \nonumber
\end{equation}
%
%
Of these, only the $E^\prime$ and $A_2^{\prime\prime}$ vibrations are infrared
active along the {\em a-b} planes and {\em c} axis, respectively, thus the
three modes we observe in this polarization along the {\em b} axis are the normally
infrared-active $E^\prime$ modes.  The 292 and 302~cm$^{-1}$ modes shift to slightly
higher frequencies with decreasing temperature, but do not display any anomalous
behavior at $T_N$.  The 246~cm$^{-1}$ mode also increases in frequency as the
temperature is reduced; however, below $T_N$ this mode begins to decrease in
frequency with decreasing temperature.   Strong spin-phonon interactions are
expected to manifest themselves as either anomalous frequency shifts with the
onset of magnetic order, or the appearance of Fano-like asymmetric line
shapes.\cite{bray,homes95,kuzmenko01,choi03,sushkov05,kim06}  The temperature
dependence of the 246~cm$^{-1}$ mode suggests weak spin-phonon coupling.

%
%
%
\begin{table}[tb]
\caption{The vibrational parameters for the fits to infrared-active modes in the
optical conductivity of FeCrAs along the {\em b} and {\em c} axis between 295 and 6~K,
where $\omega_k$, $\gamma_k$ and $\Omega_k$ are the frequency, width and strength
of the $k$th mode.  All units are in cm$^{-1}$.}
\begin{ruledtabular}
\begin{tabular}{ccc ccc ccc}
 \multicolumn{9}{c}{$E\parallel{b}$, $E^\prime$} \\
 \multicolumn{3}{c}{295~K} &
 \multicolumn{3}{c}{130~K} &
 \multicolumn{3}{c}{6~K } \\
%
 $\omega_k$ & $\gamma_k$ & $\Omega_k$ &
 $\omega_k$ & $\gamma_k$ & $\Omega_k$ &
 $\omega_k$ & $\gamma_k$ & $\Omega_k$ \\
%
 \cline{1-3} \cline{4-6} \cline{7-9}
%
%
%
 245.8 & 27.8 & 346 &
 250.9 & 23.2 & 345 &
 248.5 & 16.7 & 342 \\
%
%
 291.6 & 14.9 & 604 &
 296.9 & 12.0 & 604 &
 297.9 & 11.2 & 603 \\
%
%
 302.0 & 18.2 & 498 &
 307.4 & 16.7 & 498 &
 309.1 & 18.0 & 496 \\

%
 & & & & & \\
 \multicolumn{9}{c}{$E\parallel{c}$, $A_2^{\prime\prime}$} \\
 \multicolumn{3}{c}{295~K} &
 \multicolumn{3}{c}{130~K} &
 \multicolumn{3}{c}{6~K } \\

 $\omega_k$ & $\gamma_k$ & $\Omega_k$ &
 $\omega_k$ & $\gamma_k$ & $\Omega_k$ &
 $\omega_k$ & $\gamma_k$ & $\Omega_k$ \\
 \cline{1-3} \cline{4-6} \cline{7-9}
%
%
 123.5 & 7.9 & 540 &
 125.9 & 7.2 & 580 &
 127.0 & 5.4 & 580 \\
%
%
 263.5 & 22.5 & 530 &
 267.2 & 17.9 & 510 &
 270.5 & 19.3 & 510 \\
 318.0 & 25.4 & 960  &
 321.4 & 19.6 & 980  &
 322.4 & 17.0 & 990 \\
\end{tabular}
\end{ruledtabular}
\label{tab:irmodes}
%
%
\end{table}

%
%
Returning to the optical conductivity, at room temperature two broad Lorentz
oscillators are observed at $\omega_{L,1} \simeq 1850$~cm$^{-1}$ and $\omega_{L,2}
\simeq 6590$~cm$^{-1}$, with broad ($1/\tau_{D,1} \simeq 900$~cm$^{-1}$) and narrow
Drude ($1/\tau_{D,2} \simeq 35$~cm$^{-1}$) components.  The majority of the free-carrier
spectral weight is associated with the broad Drude component, where the spectral weight
is defined as the area under the conductivity curve over a given interval,
$N(\omega) = \int_0^\omega  \sigma_1(\omega^\prime) d\omega^\prime$.  The ratio of
the two different spectral weights will be proportional to  $(\omega_{p,D;1}/
\omega_{p,D;2})^2 \simeq 65$.  As the temperature is lowered to just above $T_N$,
the scattering rate for the strong Drude term has increased to $1/\tau_{D,1}\simeq
1080$~cm$^{-1}$, while the scattering rate for the weaker Drude term is also somewhat
larger.  These changes are reflected in the drop in the low-frequency conductivity
and the decrease in $\sigma_{dc}$, as well as the transfer of spectral weight from low
to high frequency.  In comparison, the first Lorentz oscillator has increased in frequency
to $\omega_{L,1} \simeq 2080$~cm$^{-1}$, but more surprising is the increase in the
width to $\gamma_{L,1} \simeq 5740$~cm$^{-1}$; the high-frequency oscillator changes
relatively little.

As the temperature is reduced below $T_N$, the resistivity along this direction continues
to increase; this is reflected in the increase in the scattering rates for both Drude
terms (Table~\ref{tab:drude}) and the continued decrease in the low-frequency conductivity.
At or below $T_N$, a shoulder becomes clearly visible at about 0.12~eV in Fig.~\ref{fig:sigb}(a).
%
%
In this analysis we have assumed that this absorption is present at all temperatures;
however, it could be the case that this feature is associated with the partial gapping of
the Fermi surface below $T_N$ due to the magnetic (and possible charge) order.  We find no
evidence for this scenario because there is no effective transfer of spectral weight from below
this feature to energies above it, as would be expected for a transport gap (full or partial);
this transfer of spectral weight is observed in BaFe$_2$As$_2$ below $T_{\rm SDW} \simeq 138$~K
due to the partial gapping of the Fermi surface.\cite{hu08,akrap09}   We therefore conclude that
the carrier concentration in the bands is not changing and the constraint on $\omega_{p,D;j}$
to be valid.
It is likely that this feature is a low-lying interband transition similar to those
observed in Ba(Fe$_{1-x}$Co$_x$)$_2$As$_2$.\cite{marsik13}  The change in position
and width of the low-frequency Lorentzian oscillator is probably due to the effects of
Pauli blocking with increasing temperature (this is in accord with the observation
that the high-frequency oscillator displays relatively little temperature dependence).
The appearance of this feature below $T_N$ is thus a consequence of the depletion
of low-frequency spectral weight due to the broadening of the strong Drude component.
The non-metallic resistivity is due to Drude scattering rates that increase with decreasing
temperature.
%
%

%
%
\begin{figure}[t]
\includegraphics[width=0.85\columnwidth]{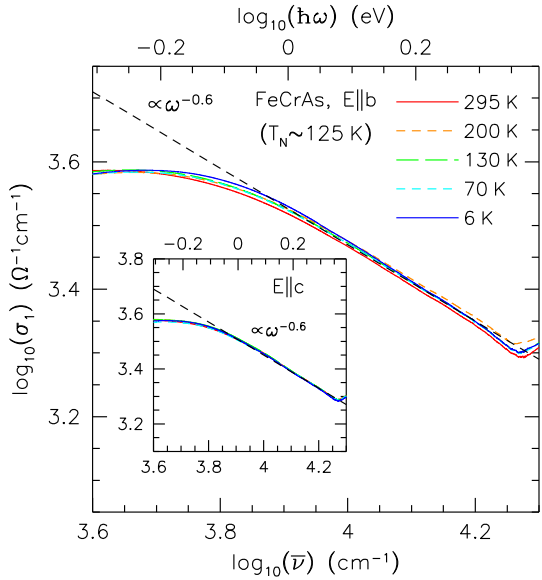}
\caption{The log-log plot of the temperature dependence of the real part of the
optical conductivity versus wavenumber ($\bar\nu=1/\lambda$) in the mid- and near-infrared
regions for FeCrAs for light polarized along the {\em b} axis.  Above about
0.7~eV the power-law behavior $\sigma_1(\omega) \propto \omega^{-0.6}$ is
clearly displayed.
Inset: The same plot for $E\parallel{c}$. }
\label{fig:loglog}
\end{figure}
%

%
In the earlier discussion of the Hund's metal, it was remarked that at high frequencies in
the incoherent regime, the optical conductivity is expected to exhibit an unusual power-law
behavior, $\sigma_1(\omega) \propto \omega^{-\alpha}$, with $0 < \alpha < 1$.  Interestingly,
the log-log plot of the real part of the optical conductivity shown in Fig.~\ref{fig:loglog}
displays just such a relationship between $6000 - 20\,000$~cm$^{-1}$, with $\sigma_1(\omega)
\propto \omega^{-0.6}$; this exponent is similar to what has been observed in this region for
a number of other pnictide and iron-chalcogenide materials.\cite{yin12}  In addition, we also
note that spectral weight is transferred from low to high frequency (i.e., above 6000~cm$^{-1}$)
as the temperature is lowered to just above $T_N$ [Fig.~\ref{fig:sigb}(a)]; this effect has
also been observed in electron- and hole-doped BaFe$_2$As$_2$ where it is argued that this
energy scale is simply the Hund's coupling $J$.\cite{wang12,schafgans12}
%
%
However, as previously discussed, the transfer of spectral weight is likely the result of the
increase of the free-carrier scattering rate(s) with decreasing temperature.  It may also be
the case that this frequency region is dominated by interband transitions described by the
two Lorentzian oscillators used to fit the conductivity, and that it is the overlapping tails
of these features that produces this fractional power law.

%
%
%
%

%
%
%
\subsection{\boldmath $E\parallel{c}$ \unboldmath}
\begin{figure}[t]
\includegraphics[width=0.85\columnwidth]{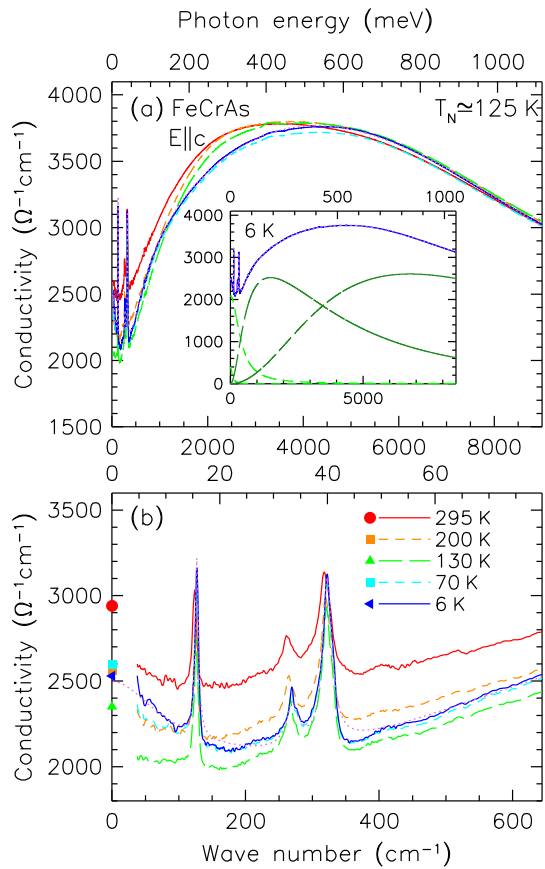}
\caption{(a) The temperature dependence of the real part of the optical conductivity
of FeCrAs for light polarized along the {\em c} axis over a wide frequency range.
The dotted line indicates the model fit to the optical conductivity at 6~K.
Inset: The decomposition of the conductivity at 6~K  into two Drude components
(dashed lines) and two Lorentzian oscillators (dash-dot lines).
(b) The temperature dependence of the real part of the optical conductivity shown
below 80~meV; the dotted line is the fit to the 6~K data.  The symbols at the origin
indicate the transport values for $\sigma_{dc}$ along the {\em c} axis.}
\label{fig:sigc}
\end{figure}

The temperature dependence of the real part of the optical conductivity for light
polarized along the {\em c} axis is shown in Fig.~\ref{fig:sigc}(a).  The low frequency
conductivity is shown in Fig.~\ref{fig:sigc}(b), where the symbols at the origin
denote the values for $\sigma_{dc}$ determined from transport measurements
(Fig.~\ref{fig:resis}).
A comparison with Fig.~\ref{fig:sigb} indicates that both the electronic and vibrational
properties of FeCrAs are anisotropic.  The same methodology that was used to fit the
optical conductivity for the {\em b} axis polarization is used for this polarization as
well.  The optical conductivity at 6~K is once again described by two Drude terms
and two Lorentz oscillators.  The Drude components consist of a broad, strong
term with $\omega_{p,D;1} \simeq 7090$~cm$^{-1}$ and $1/\tau_{D,1}\simeq 400$~cm$^{-1}$,
with a weaker component with $\omega_{p,D;2}\simeq 1450$~cm$^{-1}$ and $1/\tau_{D,2}\simeq
80$~cm$^{-1}$; however, the ratio of the strengths is now much smaller,
$(\omega_{p,D;1}/\omega_{p,D;2})^2 \simeq 25$.  The parameters for the two overdamped Lorentz
oscillators at 6~K are $\omega_{L,1}\simeq 1510$~cm$^{-1}$, $\gamma_{L,1}\simeq 4690$~cm$^{-1}$ and
$\Omega_{L,1}\simeq 26\,640$~cm$^{-1}$ and $\omega_{L,2}\simeq 6760$~cm$^{-1}$,
$\gamma_{L,2}\simeq 15\,110$~cm$^{-1}$ and $\Omega_{L,2}\simeq 48\,610$~cm$^{-1}$
(Table~\ref{tab:drude}); we also note that the strengths of the two oscillators are
more evenly matched, with $(\Omega_{L,2}/\Omega_{L,1})^2 \simeq 3.3$.

The individual contributions to the optical conductivity at 6~K are shown in the inset of
Fig.~\ref{fig:sigb}(a)  As outlined before, the fits to the optical conductivity at the
remaining temperatures have been performed by allowing only the Drude widths and the
frequencies and positions of the Lorentz oscillators to vary; the strengths are fixed.
The results are summarized in Table~\ref{tab:drude}.

%
%
%

%
%
From room temperature to just above $T_N$, the scattering rate for the strong Drude term
increases only slightly from $1/\tau_{D,1} \simeq 350$ to 440~cm$^{-1}$, while the weaker
Drude term similarly increases from $1/\tau_{D,2} \simeq 60$ to 90~cm$^{-1}$.  At the
same time, the low-frequency oscillator shifts from $\omega_{L,1}\simeq 1340$ to 1580~cm$^{-1}$,
while the damping remains more or less the same; the high-frequency oscillator shows little
temperature dependence.
%
%
As the temperature is lowered below $T_N$, the resistivity along this lattice direction
decreases before slowly increasing again at low temperature.  This is reflected in the
decrease in the scattering rate of the strong Drude component  $1/\tau_{D,1} \simeq
400$~cm$^{-1}$ at 6~K, while the scattering rate for the weak Drude term first decreases
to $1/\tau_{D2} \simeq 70$~cm$^{-1}$, before increasing slightly to $\simeq 80$~cm$^{-1}$
at 6~K.  Below $T_N$, the oscillators display little temperature dependence.  At or
below $T_N$, a weak shoulder is observed in the optical conductivity at about
0.12~eV.  Once again, there is no appreciable transfer of spectral weight from low
to high frequency associated with this feature ruling out the possibility of a transport
gap, and it is argued that this structure becomes visible as a consequence of the broadening
of the Drude terms at low temperature.

The optical conductivity in the high-frequency region again yields the unusual
$\sigma_1(\omega) \propto \omega^{-0.6}$ power-law behavior (inset of
Fig.~\ref{fig:loglog}); however, as was previously discussed, while this result is
expected for a Hund's metal in the incoherent regime, it may also be the result of
overlapping interband transitions.

%
%
The three sharp lattice modes shown in Fig.~\ref{fig:sigc}(b) are attributed to the
normally infrared-active $A_2^{\prime\prime}$ modes.  While the two modes at
$\simeq 264$ and 318~cm$^{-1}$ are similar in frequency to the $E^\prime$ modes
observed in the {\em b} polarization, the mode at $\simeq 124$~cm$^{-1}$ is
much lower in energy and also considerably narrower (Table~\ref{tab:irmodes}).
The modes increase in frequency with decreasing temperature and show
no evidence of any anomalous behavior at $T_N$.

%
%
\section{Conclusions}
The optical properties of FeCrAs reveal a weakly anisotropic material
at room temperature. Electronic structure calculations for FeCrAs show
that this system is a metal with two bands crossing $\epsilon_F$ and a
non-zero density of states; however, the temperature dependence of
the resistivity exhibits a non-metallic response.
The multiband nature of this material allows the optical properties to be fit
using a two-Drude model which yields a strong, broad term that extends well
into the mid-infrared region, and a much weaker, narrower term that is necessary
to describe the upturn in the low-frequency conductivity as well as yielding the
correct values for the dc conductivity.  The mid-infrared region may be modeled using two
bound excitations; at room temperature the strong excitation is at about $\simeq 0.8$~eV in
the planes and is almost identical along the {\em c} axis; this feature displays
little temperature dependence.  At or below $T_N$ a shoulder becomes visible in
$\sigma_1(\omega)$ at $\simeq 0.12$~eV in both polarizations; however, the absence
of any transfer of spectral weight from below to above this energy below $T_N$ suggests
that it is not associated with the opening of a partial gap and its emergence is instead
attributed to the broadening of the Drude components with decreasing temperature.
Superimposed on the optical conductivity are a number of sharp absorptions due
to the normally infrared-active $E^\prime$ and $A^{\prime\prime}_2$ modes in the
{\em a-b} planes and along the {\em c} axis, respectively.  In contrast to the modes
that are observed to increase in frequency with decreasing temperature and show no
anomalous behavior at $T_N$, the low-frequency $E^\prime$ mode at $\simeq 246$~cm$^{-1}$
softens below $T_N$, suggestive of spin-phonon coupling.
%
%
%
%

%
%
\begin{acknowledgements}
Research supported by the U.S. Department of Energy, Office of
Basic Energy Sciences, Division of Materials Sciences and Engineering
under Contract No. DE-AC02-98CH10886, Natural Science and Engineering Research
Council of Canada, and the Canadian Institute for Advanced Research.
\end{acknowledgements}
%

%
%
%

\begin{thebibliography}{51}%
\makeatletter
\providecommand \@ifxundefined [1]{%
 \@ifx{#1\undefined}
}%
\providecommand \@ifnum [1]{%
 \ifnum #1\expandafter \@firstoftwo
 \else \expandafter \@secondoftwo
 \fi
}%
\providecommand \@ifx [1]{%
 \ifx #1\expandafter \@firstoftwo
 \else \expandafter \@secondoftwo
 \fi
}%
\providecommand \natexlab [1]{#1}%
\providecommand \enquote  [1]{``#1''}%
\providecommand \bibnamefont  [1]{#1}%
\providecommand \bibfnamefont [1]{#1}%
\providecommand \citenamefont [1]{#1}%
\providecommand \href@noop [0]{\@secondoftwo}%
\providecommand \href [0]{\begingroup \@sanitize@url \@href}%
\providecommand \@href[1]{\@@startlink{#1}\@@href}%
\providecommand \@@href[1]{\endgroup#1\@@endlink}%
\providecommand \@sanitize@url [0]{\catcode `\\12\catcode `\$12\catcode
  `\&12\catcode `\#12\catcode `\^12\catcode `\_12\catcode `\%12\relax}%
\providecommand \@@startlink[1]{}%
\providecommand \@@endlink[0]{}%
\providecommand \url  [0]{\begingroup\@sanitize@url \@url }%
\providecommand \@url [1]{\endgroup\@href {#1}{\urlprefix }}%
\providecommand \urlprefix  [0]{URL }%
\providecommand \Eprint [0]{\href }%
\providecommand \doibase [0]{http://dx.doi.org/}%
\providecommand \selectlanguage [0]{\@gobble}%
\providecommand \bibinfo  [0]{\@secondoftwo}%
\providecommand \bibfield  [0]{\@secondoftwo}%
\providecommand \translation [1]{[#1]}%
\providecommand \BibitemOpen [0]{}%
\providecommand \bibitemStop [0]{}%
\providecommand \bibitemNoStop [0]{.\EOS\space}%
\providecommand \EOS [0]{\spacefactor3000\relax}%
\providecommand \BibitemShut  [1]{\csname bibitem#1\endcsname}%
\let\auto@bib@innerbib\@empty
\bibitem [{\citenamefont {Ando}\ \emph {et~al.}(1995)\citenamefont {Ando},
  \citenamefont {Boebinger}, \citenamefont {Passner}, \citenamefont {Kimura},\
  and\ \citenamefont {Kishio}}]{ando95}%
  \BibitemOpen
  \bibfield  {author} {\bibinfo {author} {\bibfnamefont {Y.}~\bibnamefont
  {Ando}}, \bibinfo {author} {\bibfnamefont {G.~S.}\ \bibnamefont {Boebinger}},
  \bibinfo {author} {\bibfnamefont {A.}~\bibnamefont {Passner}}, \bibinfo
  {author} {\bibfnamefont {T.}~\bibnamefont {Kimura}}, \ and\ \bibinfo {author}
  {\bibfnamefont {K.}~\bibnamefont {Kishio}},\ }\href {\doibase
  10.1103/PhysRevLett.75.4662} {\bibfield  {journal} {\bibinfo  {journal}
  {Phys.\ Rev.\ Lett.}\ }\textbf {\bibinfo {volume} {75}},\ \bibinfo {pages}
  {4662} (\bibinfo {year} {1995})}\BibitemShut {NoStop}%
\bibitem [{\citenamefont {Dobrosavljevic}(2012)}]{dobrosavljevic12}%
  \BibitemOpen
  \bibfield  {author} {\bibinfo {author} {\bibfnamefont {V.}~\bibnamefont
  {Dobrosavljevic}},\ }in\ \href@noop {} {\emph {\bibinfo {booktitle}
  {{Conductor Insulator Quantum Phase Transitions}}}},\ \bibinfo {editor}
  {edited by\ \bibinfo {editor} {\bibfnamefont {V.}~\bibnamefont
  {Dobrosavljevic}}\ and\ \bibinfo {editor} {\bibfnamefont {N.}~\bibnamefont
  {Trivedi}}}\ (\bibinfo  {publisher} {Oxford University Press},\ \bibinfo
  {address} {Oxford, UK},\ \bibinfo {year} {2012})\BibitemShut {NoStop}%
\bibitem [{\citenamefont {Maple}\ \emph {et~al.}(1995)\citenamefont {Maple},
  \citenamefont {Deandrade}, \citenamefont {Herrmann}, \citenamefont
  {Dalichaouch}, \citenamefont {Gajewski}, \citenamefont {Seaman},
  \citenamefont {Chau}, \citenamefont {Movshovich}, \citenamefont {Aronson},\
  and\ \citenamefont {Osborn}}]{maple95}%
  \BibitemOpen
  \bibfield  {author} {\bibinfo {author} {\bibfnamefont {M.~B.}\ \bibnamefont
  {Maple}}, \bibinfo {author} {\bibfnamefont {M.~C.}\ \bibnamefont
  {Deandrade}}, \bibinfo {author} {\bibfnamefont {J.}~\bibnamefont {Herrmann}},
  \bibinfo {author} {\bibfnamefont {Y.}~\bibnamefont {Dalichaouch}}, \bibinfo
  {author} {\bibfnamefont {D.~A.}\ \bibnamefont {Gajewski}}, \bibinfo {author}
  {\bibfnamefont {C.~L.}\ \bibnamefont {Seaman}}, \bibinfo {author}
  {\bibfnamefont {R.}~\bibnamefont {Chau}}, \bibinfo {author} {\bibfnamefont
  {R.}~\bibnamefont {Movshovich}}, \bibinfo {author} {\bibfnamefont {M.~C.}\
  \bibnamefont {Aronson}}, \ and\ \bibinfo {author} {\bibfnamefont
  {R.}~\bibnamefont {Osborn}},\ }\href {\doibase 10.1007/BF00752290} {\bibfield
   {journal} {\bibinfo  {journal} {J.\ Low.\ Temp.\ Phys.}\ }\textbf {\bibinfo
  {volume} {99}},\ \bibinfo {pages} {223} (\bibinfo {year} {1995})},\ \bibinfo
  {note} {international Euroconference on Magnetic Correlations,
  Metal-Insulator-Transitions, and Superconductivity in Novel Materials, Univ.
  Wurzburg, Wurzburg, Germany, SEP 26-30, 1994}\BibitemShut {NoStop}%
\bibitem [{\citenamefont {Mott}(1970)}]{mott70}%
  \BibitemOpen
  \bibfield  {author} {\bibinfo {author} {\bibfnamefont {N.~F.}\ \bibnamefont
  {Mott}},\ }\href {\doibase 10.1080/14786437008228147} {\bibfield  {journal}
  {\bibinfo  {journal} {Phil.\ Mag.}\ }\textbf {\bibinfo {volume} {22}},\
  \bibinfo {pages} {7} (\bibinfo {year} {1970})}\BibitemShut {NoStop}%
\bibitem [{\citenamefont {{Rosenbaum, R. and Mi, S. and Grushko, B. and
  Przepi\'ozyn\'nski, B.}}(2007)}]{rosenbaum07}%
  \BibitemOpen
  \bibfield  {author} {\bibinfo {author} {\bibnamefont {{Rosenbaum, R. and Mi,
  S. and Grushko, B. and Przepi\'ozyn\'nski, B.}}},\ }\href {\doibase
  10.1007/s10909-007-9509-4} {\bibfield  {journal} {\bibinfo  {journal} {J.\
  Low.\ Temp.\ Phys.}\ }\textbf {\bibinfo {volume} {149}},\ \bibinfo {pages}
  {314} (\bibinfo {year} {2007})}\BibitemShut {NoStop}%
\bibitem [{\citenamefont {Stewart}(2011)}]{stewart11}%
  \BibitemOpen
  \bibfield  {author} {\bibinfo {author} {\bibfnamefont {G.~R.}\ \bibnamefont
  {Stewart}},\ }\href {\doibase 10.1103/RevModPhys.83.1589} {\bibfield
  {journal} {\bibinfo  {journal} {Rev.\ Mod.\ Phys.}\ }\textbf {\bibinfo
  {volume} {83}},\ \bibinfo {pages} {1589} (\bibinfo {year}
  {2011})}\BibitemShut {NoStop}%
\bibitem [{\citenamefont {Wu}\ \emph {et~al.}(2009)\citenamefont {Wu},
  \citenamefont {McCollam}, \citenamefont {Swainson}, \citenamefont {Rourke},
  \citenamefont {Rancourt},\ and\ \citenamefont {Julian}}]{wu09}%
  \BibitemOpen
  \bibfield  {author} {\bibinfo {author} {\bibfnamefont {W.}~\bibnamefont
  {Wu}}, \bibinfo {author} {\bibfnamefont {A.}~\bibnamefont {McCollam}},
  \bibinfo {author} {\bibfnamefont {I.}~\bibnamefont {Swainson}}, \bibinfo
  {author} {\bibfnamefont {P.~M.~C.}\ \bibnamefont {Rourke}}, \bibinfo {author}
  {\bibfnamefont {D.~G.}\ \bibnamefont {Rancourt}}, \ and\ \bibinfo {author}
  {\bibfnamefont {S.~R.}\ \bibnamefont {Julian}},\ }\href {\doibase
  10.1209/0295-5075/85/17009} {\bibfield  {journal} {\bibinfo  {journal} {EPL}\
  }\textbf {\bibinfo {volume} {85}},\ \bibinfo {pages} {17009} (\bibinfo {year}
  {2009})}\BibitemShut {NoStop}%
\bibitem [{\citenamefont {Nozi\`eres}(1974)}]{nozieres74}%
  \BibitemOpen
  \bibfield  {author} {\bibinfo {author} {\bibfnamefont {P.}~\bibnamefont
  {Nozi\`eres}},\ }\href@noop {} {\bibfield  {journal} {\bibinfo  {journal}
  {J.\ Low Temp.\ Phys.}\ }\textbf {\bibinfo {volume} {17}},\ \bibinfo {pages}
  {31} (\bibinfo {year} {1974})}\BibitemShut {NoStop}%
\bibitem [{\citenamefont {Swainson}\ \emph {et~al.}(2010)\citenamefont
  {Swainson}, \citenamefont {Wu}, \citenamefont {McCollam},\ and\ \citenamefont
  {Julian}}]{swainson10}%
  \BibitemOpen
  \bibfield  {author} {\bibinfo {author} {\bibfnamefont {I.~P.}\ \bibnamefont
  {Swainson}}, \bibinfo {author} {\bibfnamefont {W.}~\bibnamefont {Wu}},
  \bibinfo {author} {\bibfnamefont {A.}~\bibnamefont {McCollam}}, \ and\
  \bibinfo {author} {\bibfnamefont {S.~R.}\ \bibnamefont {Julian}},\ }\href
  {\doibase 10.1139/P09-050} {\bibfield  {journal} {\bibinfo  {journal} {Can.
  J. Phys.}\ }\textbf {\bibinfo {volume} {88}},\ \bibinfo {pages} {701}
  (\bibinfo {year} {2010})},\ \bibinfo {note} {special issue on Neutron
  Scattering in Canada.}\BibitemShut {Stop}%
\bibitem [{\citenamefont {Tafti}\ \emph {et~al.}(2013)\citenamefont {Tafti},
  \citenamefont {Wu},\ and\ \citenamefont {Julian}}]{tafti13}%
  \BibitemOpen
  \bibfield  {author} {\bibinfo {author} {\bibfnamefont {F.~F.}\ \bibnamefont
  {Tafti}}, \bibinfo {author} {\bibfnamefont {W.}~\bibnamefont {Wu}}, \ and\
  \bibinfo {author} {\bibfnamefont {S.~R.}\ \bibnamefont {Julian}},\
  }\href@noop {} {} (\bibinfo {year} {2013}),\ \Eprint
  {http://arxiv.org/abs/arXiv:1302.4791 (unpublished)} {arXiv:1302.4791
  (unpublished)} \BibitemShut {NoStop}%
\bibitem [{\citenamefont {Rau}\ and\ \citenamefont {Kee}(2011)}]{rau11}%
  \BibitemOpen
  \bibfield  {author} {\bibinfo {author} {\bibfnamefont {J.~G.}\ \bibnamefont
  {Rau}}\ and\ \bibinfo {author} {\bibfnamefont {H.-Y.}\ \bibnamefont {Kee}},\
  }\href {\doibase 10.1103/PhysRevB.84.104448} {\bibfield  {journal} {\bibinfo
  {journal} {Phys. Rev. B}\ }\textbf {\bibinfo {volume} {84}},\ \bibinfo
  {pages} {104448} (\bibinfo {year} {2011})}\BibitemShut {NoStop}%
\bibitem [{\citenamefont {Nevidomskyy}\ and\ \citenamefont
  {Coleman}(2009)}]{nevidomskyy09}%
  \BibitemOpen
  \bibfield  {author} {\bibinfo {author} {\bibfnamefont {A.~H.}\ \bibnamefont
  {Nevidomskyy}}\ and\ \bibinfo {author} {\bibfnamefont {P.}~\bibnamefont
  {Coleman}},\ }\href {\doibase 10.1103/PhysRevLett.103.147205} {\bibfield
  {journal} {\bibinfo  {journal} {Phys. Rev. Lett.}\ }\textbf {\bibinfo
  {volume} {103}},\ \bibinfo {pages} {147205} (\bibinfo {year}
  {2009})}\BibitemShut {NoStop}%
\bibitem [{\citenamefont {Craco}\ and\ \citenamefont {Leoni}(2010)}]{craco10}%
  \BibitemOpen
  \bibfield  {author} {\bibinfo {author} {\bibfnamefont {L.}~\bibnamefont
  {Craco}}\ and\ \bibinfo {author} {\bibfnamefont {S.}~\bibnamefont {Leoni}},\
  }\href {\doibase 10.1209/0295-5075/92/67003} {\bibfield  {journal} {\bibinfo
  {journal} {EPL}\ }\textbf {\bibinfo {volume} {92}},\ \bibinfo {pages} {67003}
  (\bibinfo {year} {2010})}\BibitemShut {NoStop}%
\bibitem [{\citenamefont {Yin}\ \emph {et~al.}(2011)\citenamefont {Yin},
  \citenamefont {Haule},\ and\ \citenamefont {Kotliar}}]{yin11}%
  \BibitemOpen
  \bibfield  {author} {\bibinfo {author} {\bibfnamefont {Z.~P.}\ \bibnamefont
  {Yin}}, \bibinfo {author} {\bibfnamefont {K.}~\bibnamefont {Haule}}, \ and\
  \bibinfo {author} {\bibfnamefont {G.}~\bibnamefont {Kotliar}},\ }\href
  {\doibase 10.1038/nmat3120} {\bibfield  {journal} {\bibinfo  {journal} {Nat.
  Mater.}\ }\textbf {\bibinfo {volume} {10}},\ \bibinfo {pages} {932} (\bibinfo
  {year} {2011})}\BibitemShut {NoStop}%
\bibitem [{\citenamefont {Haule}\ \emph {et~al.}(2008)\citenamefont {Haule},
  \citenamefont {Shim},\ and\ \citenamefont {Kotliar}}]{haule08}%
  \BibitemOpen
  \bibfield  {author} {\bibinfo {author} {\bibfnamefont {K.}~\bibnamefont
  {Haule}}, \bibinfo {author} {\bibfnamefont {J.~H.}\ \bibnamefont {Shim}}, \
  and\ \bibinfo {author} {\bibfnamefont {G.}~\bibnamefont {Kotliar}},\ }\href
  {\doibase 10.1103/PhysRevLett.100.226402} {\bibfield  {journal} {\bibinfo
  {journal} {Phys. Rev. Lett.}\ }\textbf {\bibinfo {volume} {100}},\ \bibinfo
  {pages} {226402} (\bibinfo {year} {2008})}\BibitemShut {NoStop}%
\bibitem [{\citenamefont {Haule}\ and\ \citenamefont
  {Kotliar}(2009)}]{haule09}%
  \BibitemOpen
  \bibfield  {author} {\bibinfo {author} {\bibfnamefont {K.}~\bibnamefont
  {Haule}}\ and\ \bibinfo {author} {\bibfnamefont {G.}~\bibnamefont
  {Kotliar}},\ }\href {\doibase 10.1088/1367-2630/11/2/025021} {\bibfield
  {journal} {\bibinfo  {journal} {New J. Phys.}\ }\textbf {\bibinfo {volume}
  {11}},\ \bibinfo {pages} {025021} (\bibinfo {year} {2009})}\BibitemShut
  {NoStop}%
\bibitem [{\citenamefont {Okada}\ and\ \citenamefont {Yosida}(1973)}]{okada73}%
  \BibitemOpen
  \bibfield  {author} {\bibinfo {author} {\bibfnamefont {I.}~\bibnamefont
  {Okada}}\ and\ \bibinfo {author} {\bibfnamefont {K.}~\bibnamefont {Yosida}},\
  }\href {\doibase 10.1143/PTP.49.1483} {\bibfield  {journal} {\bibinfo
  {journal} {Prog. of Theor. Phys.}\ }\textbf {\bibinfo {volume} {49}},\
  \bibinfo {pages} {1483} (\bibinfo {year} {1973})}\BibitemShut {NoStop}%
\bibitem [{\citenamefont {Yin}\ \emph {et~al.}(2012)\citenamefont {Yin},
  \citenamefont {Haule},\ and\ \citenamefont {Kotliar}}]{yin12}%
  \BibitemOpen
  \bibfield  {author} {\bibinfo {author} {\bibfnamefont {Z.~P.}\ \bibnamefont
  {Yin}}, \bibinfo {author} {\bibfnamefont {K.}~\bibnamefont {Haule}}, \ and\
  \bibinfo {author} {\bibfnamefont {G.}~\bibnamefont {Kotliar}},\ }\href
  {\doibase 10.1103/PhysRevB.86.195141} {\bibfield  {journal} {\bibinfo
  {journal} {Phys. Rev. B}\ }\textbf {\bibinfo {volume} {86}},\ \bibinfo
  {pages} {195141} (\bibinfo {year} {2012})}\BibitemShut {NoStop}%
\bibitem [{\citenamefont {Dressel}\ and\ \citenamefont
  {Gr{\"u}ner}(2001)}]{dressel-book}%
  \BibitemOpen
  \bibfield  {author} {\bibinfo {author} {\bibfnamefont {M.}~\bibnamefont
  {Dressel}}\ and\ \bibinfo {author} {\bibfnamefont {G.}~\bibnamefont
  {Gr{\"u}ner}},\ }\href@noop {} {\emph {\bibinfo {title} {Electrodynamics of
  Solids}}}\ (\bibinfo  {publisher} {Cambridge University Press},\ \bibinfo
  {address} {Cambridge},\ \bibinfo {year} {2001})\BibitemShut {NoStop}%
\bibitem [{\citenamefont {Graser}\ \emph {et~al.}(2010)\citenamefont {Graser},
  \citenamefont {Kemper}, \citenamefont {Maier}, \citenamefont {Cheng},
  \citenamefont {Hirschfeld},\ and\ \citenamefont {Scalapino}}]{graser10}%
  \BibitemOpen
  \bibfield  {author} {\bibinfo {author} {\bibfnamefont {S.}~\bibnamefont
  {Graser}}, \bibinfo {author} {\bibfnamefont {A.~F.}\ \bibnamefont {Kemper}},
  \bibinfo {author} {\bibfnamefont {T.~A.}\ \bibnamefont {Maier}}, \bibinfo
  {author} {\bibfnamefont {H.-P.}\ \bibnamefont {Cheng}}, \bibinfo {author}
  {\bibfnamefont {P.~J.}\ \bibnamefont {Hirschfeld}}, \ and\ \bibinfo {author}
  {\bibfnamefont {D.~J.}\ \bibnamefont {Scalapino}},\ }\href {\doibase
  10.1103/PhysRevB.81.214503} {\bibfield  {journal} {\bibinfo  {journal} {Phys.
  Rev. B}\ }\textbf {\bibinfo {volume} {81}},\ \bibinfo {pages} {214503}
  (\bibinfo {year} {2010})}\BibitemShut {NoStop}%
\bibitem [{\citenamefont {Johnston}(2010)}]{johnston10}%
  \BibitemOpen
  \bibfield  {author} {\bibinfo {author} {\bibfnamefont {D.~C.}\ \bibnamefont
  {Johnston}},\ }\href {\doibase 10.1080/00018732.2010.513480} {\bibfield
  {journal} {\bibinfo  {journal} {Adv. Phys.}\ }\textbf {\bibinfo {volume}
  {59}},\ \bibinfo {pages} {803} (\bibinfo {year} {2010})}\BibitemShut
  {NoStop}%
\bibitem [{\citenamefont {Rotter}\ \emph {et~al.}(2008)\citenamefont {Rotter},
  \citenamefont {Tegel}, \citenamefont {Johrendt}, \citenamefont
  {Schellenberg}, \citenamefont {Hermes},\ and\ \citenamefont
  {P\"ottgen}}]{rotter08}%
  \BibitemOpen
  \bibfield  {author} {\bibinfo {author} {\bibfnamefont {M.}~\bibnamefont
  {Rotter}}, \bibinfo {author} {\bibfnamefont {M.}~\bibnamefont {Tegel}},
  \bibinfo {author} {\bibfnamefont {D.}~\bibnamefont {Johrendt}}, \bibinfo
  {author} {\bibfnamefont {I.}~\bibnamefont {Schellenberg}}, \bibinfo {author}
  {\bibfnamefont {W.}~\bibnamefont {Hermes}}, \ and\ \bibinfo {author}
  {\bibfnamefont {R.}~\bibnamefont {P\"ottgen}},\ }\href {\doibase
  10.1103/PhysRevB.78.020503} {\bibfield  {journal} {\bibinfo  {journal} {Phys.
  Rev. B}\ }\textbf {\bibinfo {volume} {78}},\ \bibinfo {pages} {020503(R)}
  (\bibinfo {year} {2008})}\BibitemShut {NoStop}%
\bibitem [{\citenamefont {Tanatar}\ \emph {et~al.}(2009)\citenamefont
  {Tanatar}, \citenamefont {Ni}, \citenamefont {Samolyuk}, \citenamefont
  {Bud'ko}, \citenamefont {Canfield},\ and\ \citenamefont
  {Prozorov}}]{tanatar09}%
  \BibitemOpen
  \bibfield  {author} {\bibinfo {author} {\bibfnamefont {M.~A.}\ \bibnamefont
  {Tanatar}}, \bibinfo {author} {\bibfnamefont {N.}~\bibnamefont {Ni}},
  \bibinfo {author} {\bibfnamefont {G.~D.}\ \bibnamefont {Samolyuk}}, \bibinfo
  {author} {\bibfnamefont {S.~L.}\ \bibnamefont {Bud'ko}}, \bibinfo {author}
  {\bibfnamefont {P.~C.}\ \bibnamefont {Canfield}}, \ and\ \bibinfo {author}
  {\bibfnamefont {R.}~\bibnamefont {Prozorov}},\ }\href {\doibase
  10.1103/PhysRevB.79.134528} {\bibfield  {journal} {\bibinfo  {journal} {Phys.
  Rev. B}\ }\textbf {\bibinfo {volume} {79}},\ \bibinfo {pages} {134528}
  (\bibinfo {year} {2009})}\BibitemShut {NoStop}%
\bibitem [{\citenamefont {Richard}\ \emph {et~al.}(2010)\citenamefont
  {Richard}, \citenamefont {Nakayama}, \citenamefont {Sato}, \citenamefont
  {Neupane}, \citenamefont {Xu}, \citenamefont {Bowen}, \citenamefont {Chen},
  \citenamefont {Luo}, \citenamefont {Wang}, \citenamefont {Dai}, \citenamefont
  {Fang}, \citenamefont {Ding},\ and\ \citenamefont {Takahashi}}]{richard10}%
  \BibitemOpen
  \bibfield  {author} {\bibinfo {author} {\bibfnamefont {P.}~\bibnamefont
  {Richard}}, \bibinfo {author} {\bibfnamefont {K.}~\bibnamefont {Nakayama}},
  \bibinfo {author} {\bibfnamefont {T.}~\bibnamefont {Sato}}, \bibinfo {author}
  {\bibfnamefont {M.}~\bibnamefont {Neupane}}, \bibinfo {author} {\bibfnamefont
  {Y.-M.}\ \bibnamefont {Xu}}, \bibinfo {author} {\bibfnamefont {J.~H.}\
  \bibnamefont {Bowen}}, \bibinfo {author} {\bibfnamefont {G.~F.}\ \bibnamefont
  {Chen}}, \bibinfo {author} {\bibfnamefont {J.~L.}\ \bibnamefont {Luo}},
  \bibinfo {author} {\bibfnamefont {N.~L.}\ \bibnamefont {Wang}}, \bibinfo
  {author} {\bibfnamefont {X.}~\bibnamefont {Dai}}, \bibinfo {author}
  {\bibfnamefont {Z.}~\bibnamefont {Fang}}, \bibinfo {author} {\bibfnamefont
  {H.}~\bibnamefont {Ding}}, \ and\ \bibinfo {author} {\bibfnamefont
  {T.}~\bibnamefont {Takahashi}},\ }\href {\doibase
  10.1103/PhysRevLett.104.137001} {\bibfield  {journal} {\bibinfo  {journal}
  {Phys. Rev. Lett.}\ }\textbf {\bibinfo {volume} {104}},\ \bibinfo {pages}
  {137001} (\bibinfo {year} {2010})}\BibitemShut {NoStop}%
\bibitem [{\citenamefont {Hu}\ \emph {et~al.}(2008)\citenamefont {Hu},
  \citenamefont {Dong}, \citenamefont {Li}, \citenamefont {Li}, \citenamefont
  {Zheng}, \citenamefont {Chen}, \citenamefont {Luo},\ and\ \citenamefont
  {Wang}}]{hu08}%
  \BibitemOpen
  \bibfield  {author} {\bibinfo {author} {\bibfnamefont {W.~Z.}\ \bibnamefont
  {Hu}}, \bibinfo {author} {\bibfnamefont {J.}~\bibnamefont {Dong}}, \bibinfo
  {author} {\bibfnamefont {G.}~\bibnamefont {Li}}, \bibinfo {author}
  {\bibfnamefont {Z.}~\bibnamefont {Li}}, \bibinfo {author} {\bibfnamefont
  {P.}~\bibnamefont {Zheng}}, \bibinfo {author} {\bibfnamefont {G.~F.}\
  \bibnamefont {Chen}}, \bibinfo {author} {\bibfnamefont {J.~L.}\ \bibnamefont
  {Luo}}, \ and\ \bibinfo {author} {\bibfnamefont {N.~L.}\ \bibnamefont
  {Wang}},\ }\href {\doibase 10.1103/PhysRevLett.101.257005} {\bibfield
  {journal} {\bibinfo  {journal} {Phys. Rev. Lett.}\ }\textbf {\bibinfo
  {volume} {101}},\ \bibinfo {pages} {257005} (\bibinfo {year}
  {2008})}\BibitemShut {NoStop}%
\bibitem [{\citenamefont {Akrap}\ \emph {et~al.}(2009)\citenamefont {Akrap},
  \citenamefont {Tu}, \citenamefont {Li}, \citenamefont {Cao}, \citenamefont
  {Xu},\ and\ \citenamefont {Homes}}]{akrap09}%
  \BibitemOpen
  \bibfield  {author} {\bibinfo {author} {\bibfnamefont {A.}~\bibnamefont
  {Akrap}}, \bibinfo {author} {\bibfnamefont {J.~J.}\ \bibnamefont {Tu}},
  \bibinfo {author} {\bibfnamefont {L.~J.}\ \bibnamefont {Li}}, \bibinfo
  {author} {\bibfnamefont {G.~H.}\ \bibnamefont {Cao}}, \bibinfo {author}
  {\bibfnamefont {Z.~A.}\ \bibnamefont {Xu}}, \ and\ \bibinfo {author}
  {\bibfnamefont {C.~C.}\ \bibnamefont {Homes}},\ }\href {\doibase
  10.1103/PhysRevB.80.180502} {\bibfield  {journal} {\bibinfo  {journal} {Phys.
  Rev. B}\ }\textbf {\bibinfo {volume} {80}},\ \bibinfo {pages} {180502}
  (\bibinfo {year} {2009})}\BibitemShut {NoStop}%
\bibitem [{\citenamefont {Homes}\ \emph {et~al.}(1991)\citenamefont {Homes},
  \citenamefont {Timusk}, \citenamefont {Wu}, \citenamefont {Altounian},
  \citenamefont {Sahnoune},\ and\ \citenamefont {Str\"om-Olsen}}]{homes91}%
  \BibitemOpen
  \bibfield  {author} {\bibinfo {author} {\bibfnamefont {C.~C.}\ \bibnamefont
  {Homes}}, \bibinfo {author} {\bibfnamefont {T.}~\bibnamefont {Timusk}},
  \bibinfo {author} {\bibfnamefont {X.}~\bibnamefont {Wu}}, \bibinfo {author}
  {\bibfnamefont {Z.}~\bibnamefont {Altounian}}, \bibinfo {author}
  {\bibfnamefont {A.}~\bibnamefont {Sahnoune}}, \ and\ \bibinfo {author}
  {\bibfnamefont {J.~O.}\ \bibnamefont {Str\"om-Olsen}},\ }\href {\doibase
  10.1103/PhysRevLett.67.2694} {\bibfield  {journal} {\bibinfo  {journal}
  {Phys. Rev. Lett.}\ }\textbf {\bibinfo {volume} {67}},\ \bibinfo {pages}
  {2694} (\bibinfo {year} {1991})}\BibitemShut {NoStop}%
\bibitem [{\citenamefont {Timusk}\ \emph {et~al.}(2013)\citenamefont {Timusk},
  \citenamefont {Carbotte}, \citenamefont {Homes}, \citenamefont {Basov},\ and\
  \citenamefont {Sharapov}}]{timusk13}%
  \BibitemOpen
  \bibfield  {author} {\bibinfo {author} {\bibfnamefont {T.}~\bibnamefont
  {Timusk}}, \bibinfo {author} {\bibfnamefont {J.~P.}\ \bibnamefont
  {Carbotte}}, \bibinfo {author} {\bibfnamefont {C.~C.}\ \bibnamefont {Homes}},
  \bibinfo {author} {\bibfnamefont {D.~N.}\ \bibnamefont {Basov}}, \ and\
  \bibinfo {author} {\bibfnamefont {S.~G.}\ \bibnamefont {Sharapov}},\ }\href
  {\doibase 10.1103/PhysRevB.87.235121} {\bibfield  {journal} {\bibinfo
  {journal} {Phys. Rev. B}\ }\textbf {\bibinfo {volume} {87}},\ \bibinfo
  {pages} {235121} (\bibinfo {year} {2013})}\BibitemShut {NoStop}%
\bibitem [{\citenamefont {Wu}\ \emph {et~al.}(2011)\citenamefont {Wu},
  \citenamefont {McCollam}, \citenamefont {Swainson},\ and\ \citenamefont
  {Julian}}]{wu11}%
  \BibitemOpen
  \bibfield  {author} {\bibinfo {author} {\bibfnamefont {W.}~\bibnamefont
  {Wu}}, \bibinfo {author} {\bibfnamefont {A.}~\bibnamefont {McCollam}},
  \bibinfo {author} {\bibfnamefont {I.~P.}\ \bibnamefont {Swainson}}, \ and\
  \bibinfo {author} {\bibfnamefont {S.~R.}\ \bibnamefont {Julian}},\ }in\ \href
  {\doibase 10.4028/www.scientific.net/SSP.170.276} {\emph {\bibinfo
  {booktitle} {Solid Compounds of Transition Elements}}},\ \bibinfo {series}
  {Solid State Phenomena}, Vol.\ \bibinfo {volume} {170},\ \bibinfo {editor}
  {edited by\ \bibinfo {editor} {\bibfnamefont {J.~L.}\ \bibnamefont {Bobet}},
  \bibinfo {editor} {\bibfnamefont {B.}~\bibnamefont {Chevalier}}, \ and\
  \bibinfo {editor} {\bibfnamefont {D.}~\bibnamefont {Fruchart}}}\ (\bibinfo
  {year} {2011})\ pp.\ \bibinfo {pages} {276--281},\ \bibinfo {note} {17th
  International Conference on Solid Compounds of Transition Elements, Imperial
  Hotel Convent Ctr, Annecy, France, Sept.\ 05-10, 2010}\BibitemShut {NoStop}%
\bibitem [{\citenamefont {Homes}\ \emph {et~al.}(1993)\citenamefont {Homes},
  \citenamefont {Reedyk}, \citenamefont {Crandles},\ and\ \citenamefont
  {Timusk}}]{homes93}%
  \BibitemOpen
  \bibfield  {author} {\bibinfo {author} {\bibfnamefont {C.~C.}\ \bibnamefont
  {Homes}}, \bibinfo {author} {\bibfnamefont {M.}~\bibnamefont {Reedyk}},
  \bibinfo {author} {\bibfnamefont {D.~A.}\ \bibnamefont {Crandles}}, \ and\
  \bibinfo {author} {\bibfnamefont {T.}~\bibnamefont {Timusk}},\ }\href
  {\doibase 10.1364/AO.32.002976} {\bibfield  {journal} {\bibinfo  {journal}
  {Appl. Opt.}\ }\textbf {\bibinfo {volume} {32}},\ \bibinfo {pages} {2976}
  (\bibinfo {year} {1993})}\BibitemShut {NoStop}%
\bibitem [{\citenamefont {Hollan}(1966)}]{hollan66}%
  \BibitemOpen
  \bibfield  {author} {\bibinfo {author} {\bibfnamefont {L.}~\bibnamefont
  {Hollan}},\ }\href@noop {} {\bibfield  {journal} {\bibinfo  {journal} {Ann.
  Chim. (Paris)}\ }\textbf {\bibinfo {volume} {1}},\ \bibinfo {pages} {437–}
  (\bibinfo {year} {1966})}\BibitemShut {NoStop}%
\bibitem [{\citenamefont {Nylund}\ \emph {et~al.}(1972)\citenamefont {Nylund},
  \citenamefont {Roger}, \citenamefont {S\'{e}nateur},\ and\ \citenamefont
  {Fruchart}}]{nyland72}%
  \BibitemOpen
  \bibfield  {author} {\bibinfo {author} {\bibfnamefont {M.}~\bibnamefont
  {Nylund}}, \bibinfo {author} {\bibfnamefont {M.}~\bibnamefont {Roger}},
  \bibinfo {author} {\bibfnamefont {J.}~\bibnamefont {S\'{e}nateur}}, \ and\
  \bibinfo {author} {\bibfnamefont {R.}~\bibnamefont {Fruchart}},\ }\href@noop
  {} {\bibfield  {journal} {\bibinfo  {journal} {J. Solid State Chem.}\
  }\textbf {\bibinfo {volume} {4}},\ \bibinfo {pages} {115–} (\bibinfo {year}
  {1972})}\BibitemShut {NoStop}%
\bibitem [{\citenamefont {Gu\'{e}rin}\ and\ \citenamefont
  {Sergent}(1977)}]{geurin77}%
  \BibitemOpen
  \bibfield  {author} {\bibinfo {author} {\bibfnamefont {R.}~\bibnamefont
  {Gu\'{e}rin}}\ and\ \bibinfo {author} {\bibfnamefont {M.}~\bibnamefont
  {Sergent}},\ }\href@noop {} {\bibfield  {journal} {\bibinfo  {journal}
  {Mater. Res. Bull.}\ }\textbf {\bibinfo {volume} {12}},\ \bibinfo {pages}
  {381–} (\bibinfo {year} {1977})}\BibitemShut {NoStop}%
\bibitem [{\citenamefont {Littlewood}\ and\ \citenamefont
  {Varma}(1991)}]{littlewood91}%
  \BibitemOpen
  \bibfield  {author} {\bibinfo {author} {\bibfnamefont {P.~B.}\ \bibnamefont
  {Littlewood}}\ and\ \bibinfo {author} {\bibfnamefont {C.~M.}\ \bibnamefont
  {Varma}},\ }\href {\doibase 10.1063/1.348195} {\bibfield  {journal} {\bibinfo
   {journal} {J. Appl. Phys.}\ }\textbf {\bibinfo {volume} {69}},\ \bibinfo
  {pages} {4979} (\bibinfo {year} {1991})}\BibitemShut {NoStop}%
\bibitem [{\citenamefont {Wooten}(1972)}]{wooten}%
  \BibitemOpen
  \bibfield  {author} {\bibinfo {author} {\bibfnamefont {F.}~\bibnamefont
  {Wooten}},\ }\href@noop {} {\emph {\bibinfo {title} {Optical Properties of
  Solids}}}\ (\bibinfo  {publisher} {Academic Press},\ \bibinfo {address} {New
  York},\ \bibinfo {year} {1972})\ pp.\ \bibinfo {pages} {244--250}\BibitemShut
  {NoStop}%
\bibitem [{\citenamefont {Singh}(1994)}]{singh}%
  \BibitemOpen
  \bibfield  {author} {\bibinfo {author} {\bibfnamefont {D.~J.}\ \bibnamefont
  {Singh}},\ }\href@noop {} {\emph {\bibinfo {title} {Planewaves,
  Pseudopotentials and the LAPW method}}}\ (\bibinfo  {publisher} {Kluwer
  Adademic},\ \bibinfo {address} {Boston},\ \bibinfo {year} {1994})\BibitemShut
  {NoStop}%
\bibitem [{\citenamefont {Singh}(1991)}]{singh91}%
  \BibitemOpen
  \bibfield  {author} {\bibinfo {author} {\bibfnamefont {D.}~\bibnamefont
  {Singh}},\ }\href {\doibase 10.1103/PhysRevB.43.6388} {\bibfield  {journal}
  {\bibinfo  {journal} {Phys. Rev. B}\ }\textbf {\bibinfo {volume} {43}},\
  \bibinfo {pages} {6388} (\bibinfo {year} {1991})}\BibitemShut {NoStop}%
\bibitem [{wie()}]{wien2k}%
  \BibitemOpen
  \href@noop {} {}\bibinfo {note} {P. Blaha, K. Schwarz, G.~K.~H. Madsen, D.
  Kvasnicka and J. Luitz, WIEN2k, {\it An augmented plane wave plus local
  orbitals program for calculating crystal properties} (Techn.
  Universit{\"{a}}t Wien, Austria, 2001).}\BibitemShut {Stop}%
\bibitem [{\citenamefont {Ishida}\ \emph {et~al.}(1996)\citenamefont {Ishida},
  \citenamefont {Takiguchi}, \citenamefont {Fujii},\ and\ \citenamefont
  {Asano}}]{ishida96}%
  \BibitemOpen
  \bibfield  {author} {\bibinfo {author} {\bibfnamefont {S.}~\bibnamefont
  {Ishida}}, \bibinfo {author} {\bibfnamefont {T.}~\bibnamefont {Takiguchi}},
  \bibinfo {author} {\bibfnamefont {S.}~\bibnamefont {Fujii}}, \ and\ \bibinfo
  {author} {\bibfnamefont {S.}~\bibnamefont {Asano}},\ }\href {\doibase
  10.1016/0921-4526(95)00538-2} {\bibfield  {journal} {\bibinfo  {journal}
  {Physica B}\ }\textbf {\bibinfo {volume} {217}},\ \bibinfo {pages} {87 }
  (\bibinfo {year} {1996})}\BibitemShut {NoStop}%
\bibitem [{\citenamefont {Allen}\ and\ \citenamefont
  {Mikkelsen}(1977)}]{allen77}%
  \BibitemOpen
  \bibfield  {author} {\bibinfo {author} {\bibfnamefont {J.~W.}\ \bibnamefont
  {Allen}}\ and\ \bibinfo {author} {\bibfnamefont {J.~C.}\ \bibnamefont
  {Mikkelsen}},\ }\href {\doibase 10.1103/PhysRevB.15.2952} {\bibfield
  {journal} {\bibinfo  {journal} {Phys. Rev. B}\ }\textbf {\bibinfo {volume}
  {15}},\ \bibinfo {pages} {2952} (\bibinfo {year} {1977})}\BibitemShut
  {NoStop}%
\bibitem [{\citenamefont {Puchkov}\ \emph {et~al.}(1996)\citenamefont
  {Puchkov}, \citenamefont {Basov},\ and\ \citenamefont {Timusk}}]{puchkov96}%
  \BibitemOpen
  \bibfield  {author} {\bibinfo {author} {\bibfnamefont {A.~V.}\ \bibnamefont
  {Puchkov}}, \bibinfo {author} {\bibfnamefont {D.~N.}\ \bibnamefont {Basov}},
  \ and\ \bibinfo {author} {\bibfnamefont {T.}~\bibnamefont {Timusk}},\ }\href
  {\doibase 10.1088/0953-8984/8/48/023} {\bibfield  {journal} {\bibinfo
  {journal} {J. Phys.: Condens. Matter}\ }\textbf {\bibinfo {volume} {8}},\
  \bibinfo {pages} {10049} (\bibinfo {year} {1996})}\BibitemShut {NoStop}%
\bibitem [{\citenamefont {Wu}\ \emph {et~al.}(2010)\citenamefont {Wu},
  \citenamefont {Bari\ifmmode \check{s}\else \v{s}\fi{}i\ifmmode~\acute{c}\else
  \'{c}\fi{}}, \citenamefont {Kallina}, \citenamefont {Faridian}, \citenamefont
  {Gorshunov}, \citenamefont {Drichko}, \citenamefont {Li}, \citenamefont
  {Lin}, \citenamefont {Cao}, \citenamefont {Xu}, \citenamefont {Wang},\ and\
  \citenamefont {Dressel}}]{wu10}%
  \BibitemOpen
  \bibfield  {author} {\bibinfo {author} {\bibfnamefont {D.}~\bibnamefont
  {Wu}}, \bibinfo {author} {\bibfnamefont {N.}~\bibnamefont {Bari\ifmmode
  \check{s}\else \v{s}\fi{}i\ifmmode~\acute{c}\else \'{c}\fi{}}}, \bibinfo
  {author} {\bibfnamefont {P.}~\bibnamefont {Kallina}}, \bibinfo {author}
  {\bibfnamefont {A.}~\bibnamefont {Faridian}}, \bibinfo {author}
  {\bibfnamefont {B.}~\bibnamefont {Gorshunov}}, \bibinfo {author}
  {\bibfnamefont {N.}~\bibnamefont {Drichko}}, \bibinfo {author} {\bibfnamefont
  {L.~J.}\ \bibnamefont {Li}}, \bibinfo {author} {\bibfnamefont
  {X.}~\bibnamefont {Lin}}, \bibinfo {author} {\bibfnamefont {G.~H.}\
  \bibnamefont {Cao}}, \bibinfo {author} {\bibfnamefont {Z.~A.}\ \bibnamefont
  {Xu}}, \bibinfo {author} {\bibfnamefont {N.~L.}\ \bibnamefont {Wang}}, \ and\
  \bibinfo {author} {\bibfnamefont {M.}~\bibnamefont {Dressel}},\ }\href
  {\doibase 10.1103/PhysRevB.81.100512} {\bibfield  {journal} {\bibinfo
  {journal} {Phys. Rev. B}\ }\textbf {\bibinfo {volume} {81}},\ \bibinfo
  {pages} {100512} (\bibinfo {year} {2010})}\BibitemShut {NoStop}%
\bibitem [{\citenamefont {Bray}\ \emph {et~al.}(1983)\citenamefont {Bray},
  \citenamefont {Iterrante}, \citenamefont {Jacobs},\ and\ \citenamefont
  {Bonner}}]{bray}%
  \BibitemOpen
  \bibfield  {author} {\bibinfo {author} {\bibfnamefont {J.~W.}\ \bibnamefont
  {Bray}}, \bibinfo {author} {\bibfnamefont {L.~V.}\ \bibnamefont {Iterrante}},
  \bibinfo {author} {\bibfnamefont {I.~S.}\ \bibnamefont {Jacobs}}, \ and\
  \bibinfo {author} {\bibfnamefont {J.~C.}\ \bibnamefont {Bonner}},\ }in\
  \href@noop {} {\emph {\bibinfo {booktitle} {Extended Linear Chain
  Compounds}}},\ \bibinfo {editor} {edited by\ \bibinfo {editor} {\bibfnamefont
  {J.~S.}\ \bibnamefont {Miller}}}\ (\bibinfo  {publisher} {Plenum Press},\
  \bibinfo {address} {New York},\ \bibinfo {year} {1983})\ pp.\ \bibinfo
  {pages} {353--415}\BibitemShut {NoStop}%
\bibitem [{\citenamefont {Homes}\ \emph {et~al.}(1995)\citenamefont {Homes},
  \citenamefont {Ziaei}, \citenamefont {Clayman}, \citenamefont {Irwin},\ and\
  \citenamefont {Franck}}]{homes95}%
  \BibitemOpen
  \bibfield  {author} {\bibinfo {author} {\bibfnamefont {C.~C.}\ \bibnamefont
  {Homes}}, \bibinfo {author} {\bibfnamefont {M.}~\bibnamefont {Ziaei}},
  \bibinfo {author} {\bibfnamefont {B.~P.}\ \bibnamefont {Clayman}}, \bibinfo
  {author} {\bibfnamefont {J.~C.}\ \bibnamefont {Irwin}}, \ and\ \bibinfo
  {author} {\bibfnamefont {J.~P.}\ \bibnamefont {Franck}},\ }\href {\doibase
  10.1103/PhysRevB.51.3140} {\bibfield  {journal} {\bibinfo  {journal} {Phys.
  Rev. B}\ }\textbf {\bibinfo {volume} {51}},\ \bibinfo {pages} {3140}
  (\bibinfo {year} {1995})}\BibitemShut {NoStop}%
\bibitem [{\citenamefont {Kuz'menko}\ \emph {et~al.}(2001)\citenamefont
  {Kuz'menko}, \citenamefont {van~der Marel}, \citenamefont {van Bentum},
  \citenamefont {Tishchenko}, \citenamefont {Presura},\ and\ \citenamefont
  {Bush}}]{kuzmenko01}%
  \BibitemOpen
  \bibfield  {author} {\bibinfo {author} {\bibfnamefont {A.~B.}\ \bibnamefont
  {Kuz'menko}}, \bibinfo {author} {\bibfnamefont {D.}~\bibnamefont {van~der
  Marel}}, \bibinfo {author} {\bibfnamefont {P.~J.~M.}\ \bibnamefont {van
  Bentum}}, \bibinfo {author} {\bibfnamefont {E.~A.}\ \bibnamefont
  {Tishchenko}}, \bibinfo {author} {\bibfnamefont {C.}~\bibnamefont {Presura}},
  \ and\ \bibinfo {author} {\bibfnamefont {A.~A.}\ \bibnamefont {Bush}},\
  }\href {\doibase 10.1103/PhysRevB.63.094303} {\bibfield  {journal} {\bibinfo
  {journal} {Phys. Rev. B}\ }\textbf {\bibinfo {volume} {63}},\ \bibinfo
  {pages} {094303} (\bibinfo {year} {2001})}\BibitemShut {NoStop}%
\bibitem [{\citenamefont {Choi}\ \emph {et~al.}(2003)\citenamefont {Choi},
  \citenamefont {Pashkevich}, \citenamefont {Lamonova}, \citenamefont
  {Kageyama}, \citenamefont {Ueda},\ and\ \citenamefont {Lemmens}}]{choi03}%
  \BibitemOpen
  \bibfield  {author} {\bibinfo {author} {\bibfnamefont {K.-Y.}\ \bibnamefont
  {Choi}}, \bibinfo {author} {\bibfnamefont {Y.~G.}\ \bibnamefont
  {Pashkevich}}, \bibinfo {author} {\bibfnamefont {K.~V.}\ \bibnamefont
  {Lamonova}}, \bibinfo {author} {\bibfnamefont {H.}~\bibnamefont {Kageyama}},
  \bibinfo {author} {\bibfnamefont {Y.}~\bibnamefont {Ueda}}, \ and\ \bibinfo
  {author} {\bibfnamefont {P.}~\bibnamefont {Lemmens}},\ }\href {\doibase
  10.1103/PhysRevB.68.104418} {\bibfield  {journal} {\bibinfo  {journal} {Phys.
  Rev. B}\ }\textbf {\bibinfo {volume} {68}},\ \bibinfo {pages} {104418}
  (\bibinfo {year} {2003})}\BibitemShut {NoStop}%
\bibitem [{\citenamefont {Sushkov}\ \emph {et~al.}(2005)\citenamefont
  {Sushkov}, \citenamefont {Tchernyshyov}, \citenamefont {{Ratcliff II}},
  \citenamefont {Cheong},\ and\ \citenamefont {Drew}}]{sushkov05}%
  \BibitemOpen
  \bibfield  {author} {\bibinfo {author} {\bibfnamefont {A.~B.}\ \bibnamefont
  {Sushkov}}, \bibinfo {author} {\bibfnamefont {O.}~\bibnamefont
  {Tchernyshyov}}, \bibinfo {author} {\bibfnamefont {W.}~\bibnamefont
  {{Ratcliff II}}}, \bibinfo {author} {\bibfnamefont {S.~W.}\ \bibnamefont
  {Cheong}}, \ and\ \bibinfo {author} {\bibfnamefont {H.~D.}\ \bibnamefont
  {Drew}},\ }\href {\doibase 10.1103/PhysRevLett.94.137202} {\bibfield
  {journal} {\bibinfo  {journal} {Phys. Rev. Lett.}\ }\textbf {\bibinfo
  {volume} {94}},\ \bibinfo {pages} {137202} (\bibinfo {year}
  {2005})}\BibitemShut {NoStop}%
\bibitem [{\citenamefont {Kim}\ \emph {et~al.}(2006)\citenamefont {Kim},
  \citenamefont {Jung}, \citenamefont {Park}, \citenamefont {Lee},
  \citenamefont {Drew}, \citenamefont {Cheong}, \citenamefont {Kim},\ and\
  \citenamefont {Choi}}]{kim06}%
  \BibitemOpen
  \bibfield  {author} {\bibinfo {author} {\bibfnamefont {J.}~\bibnamefont
  {Kim}}, \bibinfo {author} {\bibfnamefont {S.}~\bibnamefont {Jung}}, \bibinfo
  {author} {\bibfnamefont {M.~S.}\ \bibnamefont {Park}}, \bibinfo {author}
  {\bibfnamefont {S.-I.}\ \bibnamefont {Lee}}, \bibinfo {author} {\bibfnamefont
  {H.~D.}\ \bibnamefont {Drew}}, \bibinfo {author} {\bibfnamefont
  {H.}~\bibnamefont {Cheong}}, \bibinfo {author} {\bibfnamefont {K.~H.}\
  \bibnamefont {Kim}}, \ and\ \bibinfo {author} {\bibfnamefont {E.~J.}\
  \bibnamefont {Choi}},\ }\href {\doibase 10.1103/PhysRevB.74.052406}
  {\bibfield  {journal} {\bibinfo  {journal} {Phys. Rev. B}\ }\textbf {\bibinfo
  {volume} {74}},\ \bibinfo {pages} {052406} (\bibinfo {year}
  {2006})}\BibitemShut {NoStop}%
\bibitem [{\citenamefont {Marsik}\ \emph {et~al.}(2013)\citenamefont {Marsik},
  \citenamefont {Wang}, \citenamefont {R\"ossle}, \citenamefont {Yazdi-Rizi},
  \citenamefont {Schuster}, \citenamefont {Kim}, \citenamefont {Dubroka},
  \citenamefont {Munzar}, \citenamefont {Wolf}, \citenamefont {Chen},\ and\
  \citenamefont {Bernhard}}]{marsik13}%
  \BibitemOpen
  \bibfield  {author} {\bibinfo {author} {\bibfnamefont {P.}~\bibnamefont
  {Marsik}}, \bibinfo {author} {\bibfnamefont {C.~N.}\ \bibnamefont {Wang}},
  \bibinfo {author} {\bibfnamefont {M.}~\bibnamefont {R\"ossle}}, \bibinfo
  {author} {\bibfnamefont {M.}~\bibnamefont {Yazdi-Rizi}}, \bibinfo {author}
  {\bibfnamefont {R.}~\bibnamefont {Schuster}}, \bibinfo {author}
  {\bibfnamefont {K.~W.}\ \bibnamefont {Kim}}, \bibinfo {author} {\bibfnamefont
  {A.}~\bibnamefont {Dubroka}}, \bibinfo {author} {\bibfnamefont
  {D.}~\bibnamefont {Munzar}}, \bibinfo {author} {\bibfnamefont
  {T.}~\bibnamefont {Wolf}}, \bibinfo {author} {\bibfnamefont {X.~H.}\
  \bibnamefont {Chen}}, \ and\ \bibinfo {author} {\bibfnamefont
  {C.}~\bibnamefont {Bernhard}},\ }\href {\doibase 10.1103/PhysRevB.88.180508}
  {\bibfield  {journal} {\bibinfo  {journal} {Phys. Rev. B}\ }\textbf {\bibinfo
  {volume} {88}},\ \bibinfo {pages} {180508(R)} (\bibinfo {year}
  {2013})}\BibitemShut {NoStop}%
\bibitem [{\citenamefont {Wang}\ \emph {et~al.}(2012)\citenamefont {Wang},
  \citenamefont {Hu}, \citenamefont {Chen}, \citenamefont {Yuan}, \citenamefont
  {Li}, \citenamefont {Chen},\ and\ \citenamefont {Xiang}}]{wang12}%
  \BibitemOpen
  \bibfield  {author} {\bibinfo {author} {\bibfnamefont {N.~L.}\ \bibnamefont
  {Wang}}, \bibinfo {author} {\bibfnamefont {W.~Z.}\ \bibnamefont {Hu}},
  \bibinfo {author} {\bibfnamefont {Z.~G.}\ \bibnamefont {Chen}}, \bibinfo
  {author} {\bibfnamefont {R.~H.}\ \bibnamefont {Yuan}}, \bibinfo {author}
  {\bibfnamefont {G.}~\bibnamefont {Li}}, \bibinfo {author} {\bibfnamefont
  {G.~F.}\ \bibnamefont {Chen}}, \ and\ \bibinfo {author} {\bibfnamefont
  {T.}~\bibnamefont {Xiang}},\ }\href {\doibase 10.1088/0953-8984/24/29/294202}
  {\bibfield  {journal} {\bibinfo  {journal} {J. Phys.: Condens. Matter}\
  }\textbf {\bibinfo {volume} {24}},\ \bibinfo {pages} {294202} (\bibinfo
  {year} {2012})}\BibitemShut {NoStop}%
\bibitem [{\citenamefont {Schafgans}\ \emph {et~al.}(2012)\citenamefont
  {Schafgans}, \citenamefont {Moon}, \citenamefont {Pursley}, \citenamefont
  {LaForge}, \citenamefont {Qazilbash}, \citenamefont {Sefat}, \citenamefont
  {Mandrus}, \citenamefont {Haule}, \citenamefont {Kotliar},\ and\
  \citenamefont {Basov}}]{schafgans12}%
  \BibitemOpen
  \bibfield  {author} {\bibinfo {author} {\bibfnamefont {A.~A.}\ \bibnamefont
  {Schafgans}}, \bibinfo {author} {\bibfnamefont {S.~J.}\ \bibnamefont {Moon}},
  \bibinfo {author} {\bibfnamefont {B.~C.}\ \bibnamefont {Pursley}}, \bibinfo
  {author} {\bibfnamefont {A.~D.}\ \bibnamefont {LaForge}}, \bibinfo {author}
  {\bibfnamefont {M.~M.}\ \bibnamefont {Qazilbash}}, \bibinfo {author}
  {\bibfnamefont {A.~S.}\ \bibnamefont {Sefat}}, \bibinfo {author}
  {\bibfnamefont {D.}~\bibnamefont {Mandrus}}, \bibinfo {author} {\bibfnamefont
  {K.}~\bibnamefont {Haule}}, \bibinfo {author} {\bibfnamefont
  {G.}~\bibnamefont {Kotliar}}, \ and\ \bibinfo {author} {\bibfnamefont
  {D.~N.}\ \bibnamefont {Basov}},\ }\href {\doibase
  10.1103/PhysRevLett.108.147002} {\bibfield  {journal} {\bibinfo  {journal}
  {Phys. Rev. Lett.}\ }\textbf {\bibinfo {volume} {108}},\ \bibinfo {pages}
  {147002} (\bibinfo {year} {2012})}\BibitemShut {NoStop}%
\end{thebibliography}
%

%

\end{document}